\documentclass[twocolumn,english,pra,showpacs]{revtex4}
\pdfoutput=1
\usepackage{dcolumn}
\usepackage{bm}
\usepackage{amssymb}
\usepackage[english]{babel}
\usepackage{color}
\usepackage{bibmods}

\newcommand{\mt}{\mathrm}

\usepackage{graphicx}
\begin{document}

\title{Asymptotic Bound-state Model for Feshbach Resonances}
\author{T.G.\ Tiecke$^1$, M.R.\ Goosen$^2$, J.T.M.\ Walraven$^1$, and
S.J.J.M.F.\ Kokkelmans$^2$}
\affiliation{$^1$Van der Waals-Zeeman Institute of the University of Amsterdam, 1018 XE
The Netherlands}
\affiliation{$^2$Eindhoven University of Technology, P.O. Box 513, 5600MB Eindhoven, The
Netherlands}
\date{\today }

\begin{abstract}
We present an Asymptotic Bound-state Model which can be used to accurately
describe all Feshbach resonance positions and widths in a two-body system.
With this model we determine the coupled bound states of a particular
two-body system. The model is based on analytic properties of the two-body
Hamiltonian, and on asymptotic properties of uncoupled bound states in the
interaction potentials. In its most simple version, the only necessary
parameters are the least bound state energies and actual potentials are not
used. The complexity of the model can be stepwise increased by introducing
threshold effects, multiple vibrational levels and additional potential
parameters. The model is extensively tested on the $^6$Li-$^{40}$K system
and additional calculations on the $^{40}$K-$^{87}$Rb system are presented.
\end{abstract}

\maketitle



\section{Introduction}

\label{sect:Intro} The field of ultracold atomic gases has been rapidly
growing during the past decades. One of the main sources of growth is the
large degree of tunability to employ ultracold gases as model quantum
systems \cite{varenna08,blochdalibard08}. In particular the strength of the
two-body interaction parameter, captured by the scattering-length $a$, can
be tuned over many orders of magnitude. A quantum system can be made
repulsive ($a>0$), attractive ($a<0$), non-interacting ($a=0$) or strongly
interacting ($|a|\rightarrow \infty$) in a continuous manner by means of
Feshbach resonances~\cite{chin08}. These resonances are induced by external fields:
magnetically induced Feshbach resonances are conveniently used for
alkali-metal atoms, while optically induced Feshbach resonances seem more
promising for e.g.~alkaline-earth atoms. In this paper we
consider magnetically induced resonances only.

Feshbach resonances depend crucially on the existence of an internal atomic
structure, which can be modified by external fields. For alkali-metal atoms,
this structure is initiated by the hyperfine interaction, which can be
energetically modified by a magnetic field via the Zeeman interaction. For a
given initial spin state, its collision threshold and its two-body bound
states depend in general differently on the magnetic field. A Feshbach
resonance occurs when the threshold becomes degenerate with a bound state.
Accurate knowledge of the Feshbach resonance structure is crucial for
experiments.

The two-body system has to be solved to obtain the bound state solutions.
Since the interactions have both orbital and spin degrees of freedom, this
results in a set of radially coupled Schr\"odinger equations in the spin
basis. The set of equations is referred to as Coupled Channels equations~\cite{stoof88}, and can be solved numerically. Quite often it is far from
trivial to obtain reliable predictions for the two-body problem, due to
several reasons: the ab-initio interaction potentials are usually not
accurate enough to describe ultracold collisions. Therefore these potentials
have to be modelled by adding and modifying potential parameters. A full
calculation for all spin combinations and all potential variations is very
time-consuming. Moreover, one can easily overlook some features of the bound
state spectrum due to numerical issues such as grid sizes and numerical
accuracy. This is also due to a lack of insight of the general resonance
structure, which is often not obvious from the numerical results.

Given the above, there is certainly a need for fast and simple models to
predict and describe Feshbach resonances, which allow for a detailed insight
in the resonance structure. In the last decade various simple models have
been developed for ultracold collisions \cite{houbiers98,vogels98,hanna09},
which vary significantly in terms of complexity, accuracy and applicability.
In all these models the radial equation plays a central role in describing the
Feshbach resonances.

In this Paper we present in detail the Asymptotic Bound-state Model (ABM).
This model, briefly introduced in Ref.~\cite{wille}, and extended in Ref.~\cite{dePRL} was successfully applied to the Fermi-Fermi mixture of $^{6}$Li
and $^{40}$K. In Ref.~\cite{wille} the observed loss features were assigned
to 13 Feshbach resonances with high accuracy, and the obtained parameters
served as an input to a full coupled channels analysis. The ABM builds on an
earlier model by Moerdijk et al.~\cite{moerdijk} for homonuclear systems,
which was also applied by Stan et al.~\cite{stan} for heteronuclear systems.
This earlier model neglects the mixing of singlet and triplet states,
therefore allowing the use of uncoupled orbital and spin states. In the ABM we
make use of the radial singlet and triplet eigenstates and include the
coupling between them. This crucial improvement makes the whole approach in
principle exact, and it allows for a high degree of accuracy given a limited
number of parameters.

We show how we can systematically extend the most simple version of ABM to
predict the width of the Feshbach resonances by including threshold
behavior. Additionally we allow for the inclusion of multiple vibrational
levels and parameter for the spatial wavefunction overlap. The fact that ABM
is computationally light provides the possibility to map out the available
Feshbach resonance positions and widths for a certain system, as has been
shown in Ref. \cite{dePRL}. Throughout the paper we will use the $^{6}$Li-$^{40}$K mixture as a model system to illustrate all introduced aspects.
Additionally, we present ABM calculations on the $^{40}$K-$^{87}$Rb mixture
to demonstrate its validity on a more complex system, comparing it with
accurate coupled channel calculations \cite{juliennePrivate09}. The case of
metastable helium atoms where each atom has an electron spin of $s=1$ and
the interaction occurs through singlet, triplet and quintet interaction
potentials we discuss elsewhere \cite{heliumABM}.

In the following we describe the ABM (Sec. \ref{sect:ABM}) and various
methods to obtain the required input parameters. In Sec. \ref{sect:Application} the ABM is applied to the three physical systems and in
Sec. \ref{sect:OpenChannel} we introduce the coupling to the open channel to
predict the width of Feshbach resonances. In Section \ref{sect:Discussion}
we summarize our findings and comment on further extensions of the model.

\section{Asymptotic Bound-state Model\label{sect:ABM}}

In this section we give a detailed description of the asymptotic
bound state model. In Section \ref{sect:ABMoverview} we start with
a general overview of the model which is described in more detail
in the subsequent sections
\ref{sec:InternalEnergy} to \ref{sect:ABS}.

\subsection{Overview\label{sect:ABMoverview}}

In the ABM we consider two atoms, $\alpha $ and $\beta $, in their
electronic ground-state. To search for Feshbach resonances we use the
effective Hamiltonian \cite{tiesinga}
\begin{equation}
\mathcal{H}=\mathcal{H}^{\mathrm{rel}}+\mathcal{H}^{\mathrm{int}}.  \label{1}
\end{equation}Here $\mathcal{H}^{\mathrm{rel}}=\mathbf{p}^{2}/2\mu +\mathcal{V}$ describes
the relative motion of the atoms in the center of mass frame: the first term
is the relative kinetic energy, with $\mu $ the reduced mass, the second
term the effective interaction potential $\mathcal{V}$. The Hamiltonian $\mathcal{H}^{\mathrm{int}}$ stands for the internal energy of the two atoms.

We will represent $\mathcal{H}^{\mathrm{int}}$ by the hyperfine and Zeeman
contributions to the internal energy (Section \ref{sec:InternalEnergy}).
Therefore, $\mathcal{H}^{\mathrm{int}}$ is diagonal in the Breit-Rabi pair 
basis $\{|\alpha \beta \rangle \}$ with eigen-energies $E_{\alpha \beta }$ and
typically dependent on the magnetic field $B$. The internal states $|\alpha
\beta \rangle $ in combination with the quantum number $l$ for the angular
momentum of the relative motion define the \textit{scattering channels} $\left( \alpha \beta ,l\right) $.

Because the effective potential $\mathcal{V}$ is in general not diagonal in
the pair basis $\{|\alpha \beta \rangle \}$, the internal states of the
atoms can change in collisions. To include the coupling of the channels by $\mathcal{V}$, we transform from 
the pair basis to a spin basis $\{|\sigma \rangle \}$ in which $\mathcal{H}^{\mathrm{rel}}$ is diagonal. We
will restrict ourselves (Section \ref{sec:RelativeHamiltonian}) to effective
potentials $\mathcal{V}$ which are diagonal in $S$, the quantum number of the
total electron spin $\mathbf{S=s}_{\alpha }\mathbf{+s}_{\beta }$, where $\mathbf{s}_{\alpha }$ and $\mathbf{s}_{\beta }$ are the electron spins of
the colliding atoms. The effective potential can thus be written as $\mathcal{V}(r)=\sum_{S}|S\rangle V_{S}(r)\langle S|$, where $r$ is the
interatomic separation. The examples discussed in this paper are alkali
atoms $\left( s=1/2\right) $ which lead to a decomposition in singlet $(S=0) $ and triplet potentials $(S=1)$.

The eigenstates of $\mathcal{H}^{\mathrm{rel}}$ (bound-states and
scattering states) are solutions of the 
Schr\"{o}dinger equations for given value
of $l$, using effective potentials $V_{S}^{l}(r)$ in which the centrifugal
forces are included (Section \ref{sec:RelativeHamiltonian}). Since the effective potentials are
central interactions, a separation of variables can be performed to describe the wavefunction
as a product of a radial and angular part, $|\Psi\rangle = |\psi\rangle | Y^l_{m_l} \rangle$.
The ABM
solves the Schr\"{o}dinger equation for the
Hamiltonian (\ref{1}) starting from a restricted set of (typically just a
few) \textit{discrete} eigenstates $|\psi _{\nu}^{Sl}\rangle |Y^l_{m_l}\rangle $ of $\mathcal{H}^{\mathrm{rel}}$, using their binding 
energies $\epsilon _{\nu}^{Sl} $ as free parameters. The continuum states are not used in the model. The
set $\{|\psi _{\nu}^{Sl}\rangle \}$ corresponds to the bound-state
wavefunctions $\psi _{\nu}^{Sl}(r)=\langle r|\psi _{\nu}^{Sl}\rangle $ in
the effective potentials $V_{S}^{l}(r)$, with $\nu $ and $l$ being the
vibrational and rotational quantum numbers, respectively.

The ABM solutions are obtained by diagonalization of the Hamiltonian (\ref{1}) using the restricted set of bound states $\{|\psi _{\nu}^{Sl}\rangle
|\sigma \rangle \}$. This is equivalent to solving the secular equation
\begin{equation}
\det |(\epsilon _{\nu}^{Sl}-E_{b})\delta _{\nu l\sigma, \nu ^{\prime} l^{\prime }\sigma ^{\prime }}
+\langle \psi _{\nu}^{Sl}|\psi _{\nu^{\prime
}}^{S^{\prime }l}\rangle \langle \sigma |\mathcal{H}^{\mathrm{int}}
|\sigma ^{\prime }\rangle |=0,  \label{eq:secular}
\end{equation}
where we have used the orthonormality of $|Y^l_{m_l}\rangle$. 
The roots $E_{b}$ represent the eigenvalues of
$\mathcal{H}$ which are shifted with respect to the \textit{bare}
levels $\epsilon _{\nu}^{Sl}$ due to the presence of the coupling
term $\mathcal{H}^{\mathrm{int}}$. The roots $E_{b}$ will be
accurate as long as the influence of the continuum solutions is
small. Since the bound-state wavefunctions $\psi _{\nu}^{Sl}(r) $
are orthonormal for a given value of $S$ and $l$, the Franck-Condon
factors are $\langle \psi _{\nu}^{Sl}|\psi _{\nu^{\prime
}}^{S l}\rangle =\delta _{\nu \nu ^{\prime }}$. The eigenstates of
$\mathcal{H}$ define bound states in the system of \textit{coupled
channels}. 

We define the \textit{entrance channel} $\left( \alpha \beta ,l\right) _{0}$ by the internal states 
$|\alpha \beta \rangle $ for which we want to find Feshbach resonances with a 
given angular momentum state of $l=0,1,\cdots $.
The energy $E_{\alpha \beta }^{0}(B)$ of two free
atoms at rest in the entrance channel defines the \textit{threshold energy},
which separates the continuum of scattering states from the discrete set of
bound states. In the ABM we define $\mathcal{H}$ relative to this energy.
With this convention the threshold energy always corresponds to $E=0$,
irrespective of the magnetic field. Further, we consider only entrance
channels that are stable against spin exchange relaxation.

Since in the ABM we only consider bound states, and therefore are in the 
regime $E<0$, all channels are energetically \textit{closed}. 
Collisions in the entrance channel would have a collision energy of $E>0$ and the entrance channel would
be energetically \textit{open}, i.e. the atoms are not bound and have a finite probability of
reaching $r=\infty$ in this channel.
Although all channels are closed in the ABM we will refer 
to the entrance channel as the open channel, anticipating on the inclusion of threshold
effects in Section \ref{sect:OpenChannel}.

In the Sections \ref{sec:InternalEnergy}-\ref{sect:ABS} we discuss the ABM
in its simplest form, where level broadening by coupling to the continuum is
neglected \cite{wille}. In this approximation, Feshbach resonances are
predicted for magnetic fields $B_{0}$ where a bound level crosses the
threshold, $E_{b}=\mu _{rel}(B-B_{0})$ with $\mu _{rel}\equiv \partial
E_{b}/\partial B|_{B=B_{0}}$, and where coupling to the continuum is in
principle allowed by conservation of the angular momentum. To determine the
crossings the diagonalization (\ref{eq:secular}) has to be carried out as a
function of magnetic field.

The procedure becomes particularly simple when the coupling strength $\mathcal{H}^{\mathrm{int}}$ is small compared to the separation of the
ro-vibrational levels in the various potentials $V_{S}^{l}(r)$ because in
this case the basis set can be restricted to only the least bound level in
each of the potentials $V_{S}^{l}(r)$. In this case the set of levels $\{\epsilon _{\nu}^{Sl}\}$ reduces 
to a small number, $\{\epsilon ^{Sl}\}$, with $|s_{\alpha }-s_{\beta }|\leq S\leq s_{\alpha }+s_{\beta }$. In the
case of the alkalis only two levels, $\epsilon ^{0l}$ and $\epsilon
^{1l}$, are relevant for each value of $l$ . Further, as will be shown in
Section \ref{sect:ABS}, the least bound states have the long-range behavior
of \textit{asymptotically bound states}, which makes it possible to estimate
the value of Franck-Condon factors $\langle \psi _{\nu}^{Sl}|\psi
_{\nu^{\prime }}^{S^{\prime }l}\rangle $ without detailed
knowledge of the short-range behavior of the potentials $V_{S}^{l}(r)$. This
reduces the diagonalization (\ref{eq:secular}) to the diagonalization of a
spin Hamiltonian. Treating the $\{\epsilon ^{Sl}\}$ as fitting
parameters, their values can be determined by comparison with a hand full of
experimentally observed resonances. Once these $\{\epsilon ^{Sl}\}$ are
known, the position of all Feshbach resonances associated with these levels
can be predicted. As this procedure does not involve numerical solution of
the Schr\"{o}dinger equation for the relative motion it provides an enormous
simplification over coupled channels calculations.

In Section \ref{sect:OpenChannel} we turn to the extended version of the ABM
in which also the coupling to the open channel is taken into account. The
presence of such a coupling gives rise to a shift $\Delta$
of the uncoupled levels and above threshold to a broadening $\Gamma$ \cite{dePRL}. The width of a
Feshbach resonance is related to the lifetime $\tau =\hbar /\Gamma $ of the
bound-state above threshold and provides a measure for the coupling to the
continuum. In magnetic field units the width $\Delta B$ is related to the
scattering length by the expression \cite{moerdijk}
\begin{equation}
\label{eq:dispersiveA}
a(B)=a_{bg}\left( 1-\frac{\Delta B}{B-B_{0}}\right),
\end{equation}
where $a_{bg}$ is the background scattering length. 
Interestingly, the width $\Delta B$ can also be determined with
the same restricted basis set $\{|\psi _{\nu}^{Sl}\rangle \}$,
which does not include continuum states. 
In Section \ref{sect:OpenChannel}
this is shown for the simplest case where only a single level is
resonant and the resonance width can be found from the coupling of
two bound-state levels: the resonant level and the least-bound
level in the entrance channel.

\subsection{Internal energy\label{sec:InternalEnergy}}

\begin{figure}[ht!]
\includegraphics[width=8.3088cm]{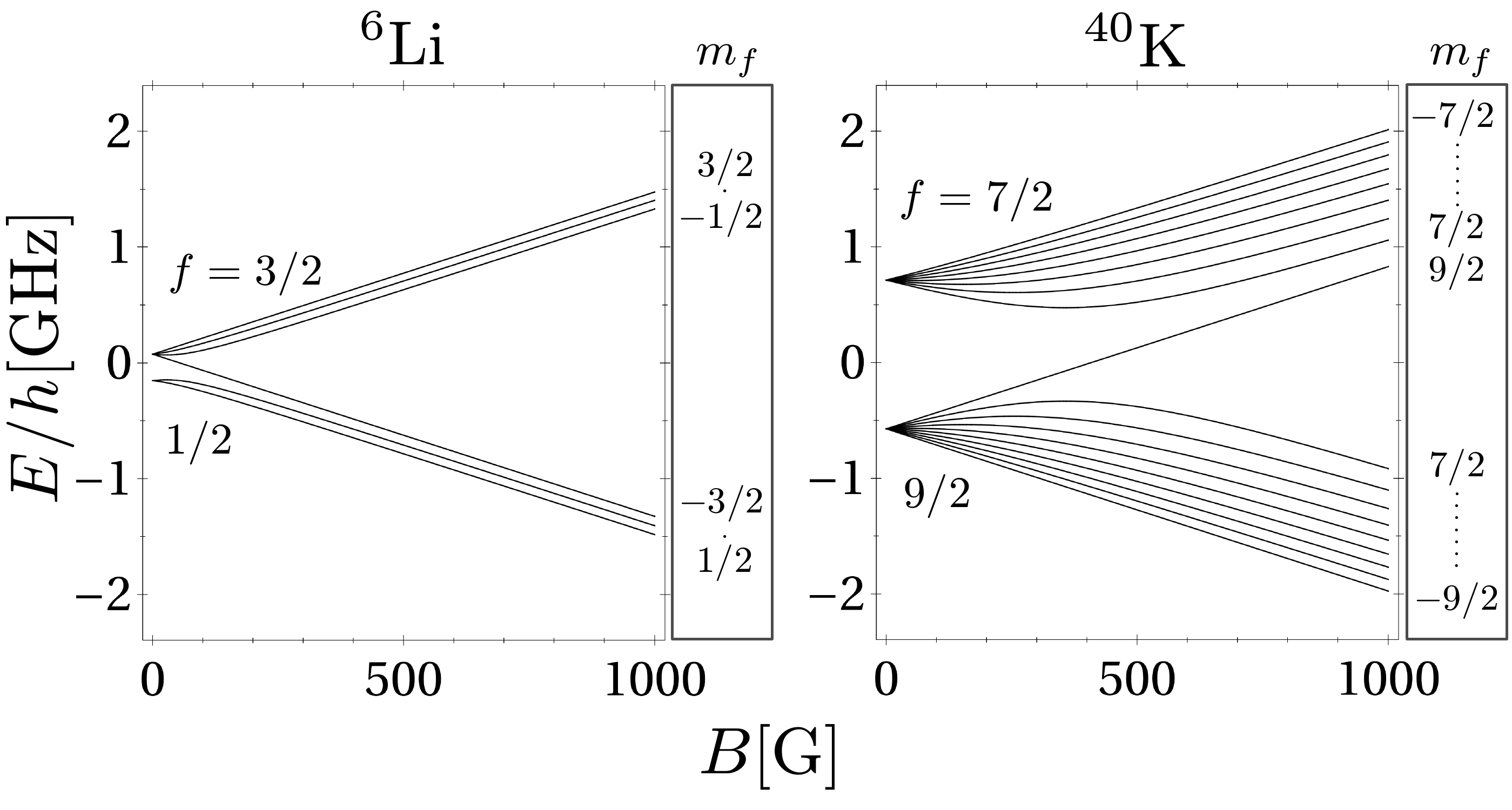}
\caption{The single atom hyperfine diagrams for $^{6}$Li and $^{40}$K. The curves 
correspond to the eigenvalues of $\mathcal{H}^{\mathrm{A}}$ and are labeled by the 
zero field quantum numbers $|fm_{f}\rangle $.} \label{fig:LiKhyperfine}
\end{figure}

To describe the internal energy of the colliding atoms we restrict the
atomic Hamiltonian to the hyperfine and Zeeman interactions
\begin{eqnarray}
\mathcal{H}^{\mathrm{A}} &=&\mathcal{H}^{\mathrm{hf}} + \mathcal{H}^{\mathrm{Z}}\\
& = & \frac{a_{\mathrm{hf}}}{\hbar ^{2}}\mathbf{i}\cdot
\mathbf{s}+(\gamma _{e}\mathbf{s}-\gamma _{i}\mathbf{i})\cdot \mathbf{B},
\end{eqnarray}
where $\mathbf{s}$ and $\mathbf{i}$ are the electron and nuclear spins
respectively, $\gamma _{e}$ and $\gamma _{i}$ are their respective
gyromagnetic ratios, $a_{\mathrm{hf}}$ is the hyperfine energy and $\mathbf{B}$ is the externally applied magnetic field. The hyperfine interaction
couples the electron and nuclear spin which add to a total angular momentum $\mathbf{f}=\mathbf{s}+\mathbf{i}$. In Fig.~\ref{fig:LiKhyperfine} the well
known Breit-Rabi diagrams of $^{6}$Li and $^{40}$K are shown, the curves
correspond to the eigenvalues of $\mathcal{H}^{\mathrm{A}}$. The one-atom
hyperfine states are labeled $|fm_{f}\rangle $, although $f$ is only a good
quantum number in the absence of an external magnetic field.

By labeling the colliding atoms with $\alpha $ and $\beta $, the two-body
internal Hamiltonian becomes $\mathcal{H}^{\mathrm{int}}=\mathcal{H}_{\mathrm{\alpha }}^{\mathrm{A}}+\mathcal{H}_{\mathrm{\beta }}^{\mathrm{A}}$ and the spin state of the colliding pair can be described in
the Breit-Rabi pair basis $|\alpha \beta \rangle
\equiv |f_{\alpha }m_{f_{\alpha }},f_{\beta }m_{f_{\beta }}\rangle
\equiv |f,m_{f}\rangle _{\alpha }\otimes |f,m_{f}\rangle _{\beta
}$. The corresponding energy of two free atoms at rest defines the
$B$-dependent threshold energy introduced in Section
\ref{sect:ABMoverview}.


\subsection{Relative Hamiltonian\label{sec:RelativeHamiltonian}}

The bound eigenstates of $\mathcal{H}^{\mathrm{rel}}$ play a central role in
the determination of the coupled bound states $\mathcal{H}$ responsible for the Feshbach 
resonances. The relative Hamiltonian includes the effective interaction $\mathcal{V}$
resulting from all Coulomb interactions between the nuclei and electrons of
both atoms \footnote{The much weaker magnetic dipole-dipole interactions are
neglected.}. This effective interaction is isotropic and depends
only on the quantum number $S$ associated with the total electron
spin. 
For these
central potentials the two-body solutions will depend on the orbital
quantum number $l$, but not on its projection $m_{l}$. In the
absence of any anisotropic interaction both $l$ and $m_{l}$ are
good quantum numbers of $\mathcal{H}^{\mathrm{rel}}$ and
$\mathcal{H}.$

We specify the ABM basis states as $\{|\psi _{\nu}^{Sl}\rangle |\sigma
\rangle \}$. Here the spin basis states $|\sigma \rangle \equiv |SM_{S}\mu
_{\alpha }\mu _{\beta }\rangle $ are determined by the spin quantum number $S $ 
and the magnetic quantum numbers $M_{S}$, $\mu _{\alpha }$, and $\mu
_{\beta }$ of the $\mathbf{S}$, $\mathbf{i}_{\alpha }$ and $\mathbf{i}_{\beta }$ 
operators, respectively. The sum $M_{F}=M_{S}+\mu _{\alpha }+\mu
_{\beta }$ is conserved by the Hamiltonian $\mathcal{H}$. This limits the
number of spin states which have to be included in the set ${|\sigma \rangle}$. The bound-state
wavefunctions $\psi _{\nu}^{Sl}(r)$ for the singlet and triplet potentials,
characterized by the vibrational and rotational quantum numbers $\nu $ and $l $, satisfy the reduced radial wave equation of $\mathcal{H}^{\mathrm{rel}}$
for specific values of $S$ and $l$,
\begin{equation}
\left[ -\frac{\hbar ^{2}}{2\mu }\frac{d^{2}}{dr^{2}}+V_{S}^{l}(r)\right]
\psi _{\nu}^{Sl}(r)=\epsilon _{\nu}^{Sl}\psi _{\nu}^{Sl}(r).
\label{eq:Hrel1D}
\end{equation}Here $V_{S}^{l}(r)\equiv V_{S}(r)+l(l+1)\hbar ^{2}/(2\mu r^{2})$
represents the interaction potentials including the
centripetal forces. The corresponding binding energies are given
by $\epsilon _{\nu}^{Sl}$. In this paper we mainly focus on
heteronuclear systems, however, the ABM works equally well for
homonuclear systems. In the latter case one would rather use a
symmetrized spin basis $|\sigma \rangle \equiv
|SM_{S}IM_{I}\rangle $, where $I$ is the total nuclear spin and
$M_{I}$ is the magnetic quantum number for $\mathbf{I} = \mathbf{i}_{\alpha} + \mathbf{i}_{\beta}$ as
described in Ref.~\cite{moerdijk}.

\subsection{Diagonalization of $\mathcal{H}$\label{sec:FullHamiltonian}}

In the ABM basis $\{|\psi _{\nu}^{Sl}\rangle |\sigma \rangle \}$ the
Zeeman term $\mathcal{H}_{\mathrm{Z}}$ is diagonal with 
\begin{equation}
E_{\sigma }^{Z} = \langle \sigma |\mathcal{H}_{\mathrm{Z}}|\sigma
\rangle=\hbar (\gamma _{e}M_{S}-\gamma _{\alpha }\mu _{\alpha }-\gamma
_{\beta }\mu _{\beta })B
\end{equation} the Zeeman energy of state $|\sigma \rangle$. 
As the orbital angular momentum is conserved, we can solve 
Eq.~(\ref{eq:secular}) separately for every $l$ subspace. Since the set $\{|\psi
_{\nu}^{Sl}\rangle |\sigma \rangle \}$ is orthonormal the secular equation
takes (for a given value of $l$) the form
\begin{equation}
\det |(\epsilon _{\nu}^{Sl}+E_{\sigma}^{Z} -E_{b})\delta _{\nu \sigma ,\nu ^{\prime} \sigma ^{\prime }}+\eta _{\nu ,\nu^{\prime }}^{S,S ^{\prime }}\langle
\sigma |\mathcal{H}_{\mathrm{hf}}|\sigma ^{\prime }\rangle |=0,
\end{equation}
where $\eta _{\nu ,\nu^{\prime }}^{S,S ^{\prime }}=\langle 
\psi _{\nu}^{Sl}|\psi _{\nu^{\prime }}^{S^{\prime } l}\rangle $ are Franck-Condon
factors between the different $S$ states, which are numbers in the
range $0\leq |\eta _{\nu ,\nu^{\prime}}^{S,S ^{\prime }}|\leq 1$ 
for $ S\neq S^{\prime } $ and $\eta _{\nu ,\nu'}^{S,S}= \delta_{\nu,\nu'}$. 
Repeating the procedure as a
function of magnetic field the energy level diagram of all bound
states in the system of coupled channels is obtained.

It is instructive to separate the hyperfine contribution into two parts, $%
\mathcal{H}_{\mathrm{hf}}=\mathcal{H}_{\mathrm{hf}}^{+}+\mathcal{H}_{\mathrm{%
hf}}^{-}$, where 
\begin{equation}
\mathcal{H}_{\mathrm{hf}}^{\pm }=\frac{a_{\mathrm{hf}}^{\alpha }}{2\hbar ^{2}%
}\left( \mathbf{s}_{\alpha }\pm \mathbf{s}_{\beta }\right) \cdot \mathbf{i}%
_{\alpha }\pm \frac{a_{\mathrm{hf}}^{\beta }}{2\hbar ^{2}}\left( \mathbf{s}%
_{\alpha }\pm \mathbf{s}_{\beta }\right) \cdot \mathbf{i}_{\beta }.
\label{eq:HF+/-}
\end{equation}%
Because $\mathcal{H}_{\mathrm{hf}}^{+}$ conserves $S$, it couples the ABM
states only within the singlet and triplet manifolds. The term $\mathcal{H}_{%
\mathrm{hf}}^{-}$ is off-diagonal in the ABM basis, hence couples singlet to triplet.
Accordingly, also the hyperfine term in the
secular equation separates into two parts 
\begin{equation}
\eta _{\nu ,\nu^{\prime }}^{S,S ^{\prime }}\langle \sigma |\mathcal{H}_{%
\mathrm{hf}}|\sigma ^{\prime }\rangle =\delta _{\nu ,\nu ^{\prime}} 
\langle \sigma |\mathcal{H}_{\mathrm{%
hf}}^{+}|\sigma ^{\prime }\rangle +\eta _{\nu ,\nu^{\prime }}^{S,S ^{\prime
}}\langle \sigma |\mathcal{H}_{\mathrm{hf}}^{-}|\sigma ^{\prime }\rangle .
\label{eq:HFseparation}
\end{equation}%
Note that the second term of Eq.\thinspace (\ref{eq:HFseparation}) is zero \textit{unless}
$S\neq S^{\prime }$. This term was neglected in the models of Refs.~%
\cite{moerdijk,stan}. This is a good approximation if no Feshbach resonances occur 
near magnetic fields where the energy difference between singlet and triplet levels is on the
order of the hyperfine energy. However, for a generic case
this term cannot be neglected. 

\begin{figure}[ht!]
\includegraphics[width=3.0441in]{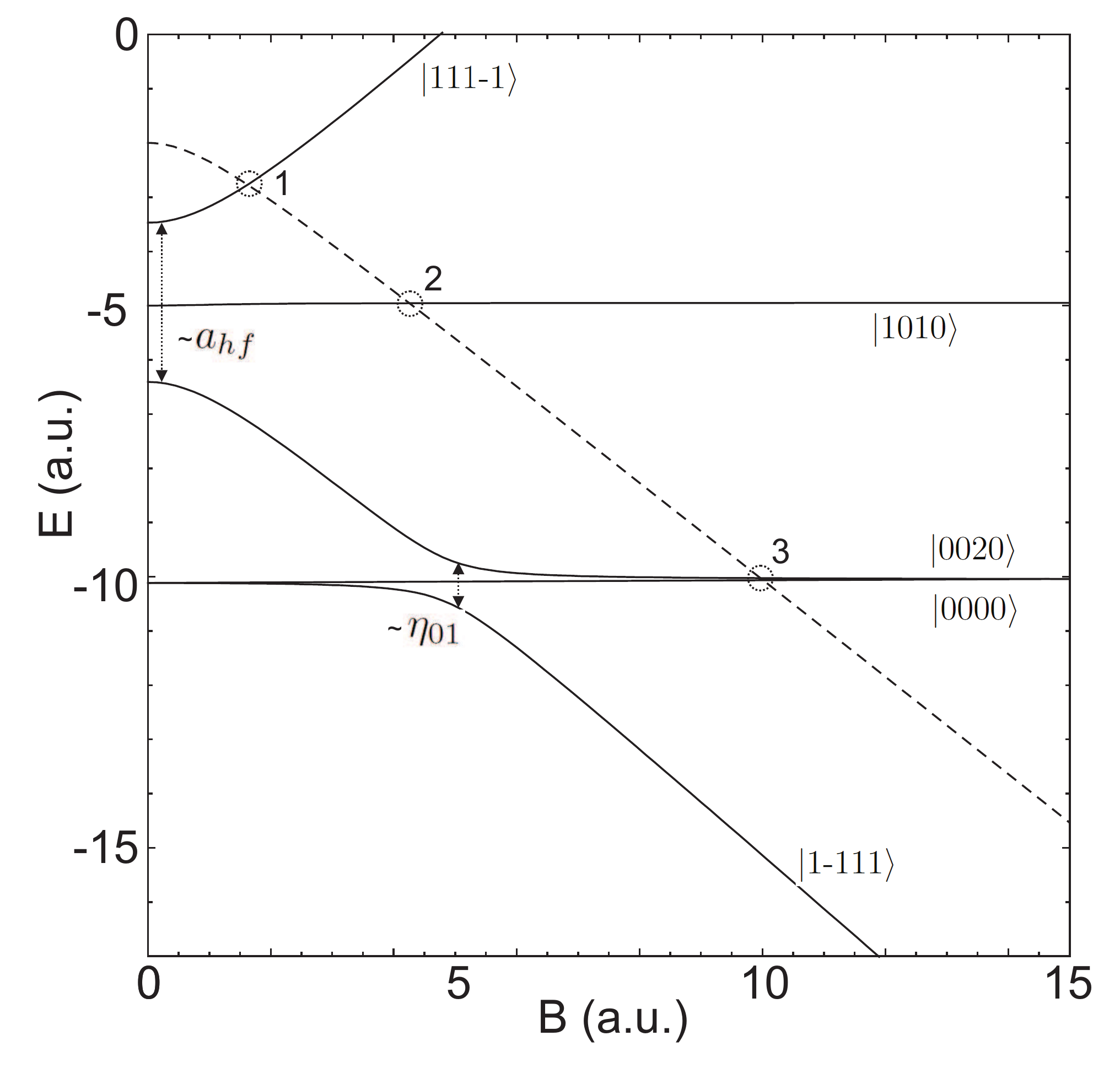}
\caption{ABM calculation for a
fictitious homonuclear system with $S=1$\ and $I=2$ for an entrance channel 
with $M_{F}=M_{S}+M_{I}=0$ and $l=0$. The threshold energy of the 
entrance channel $E_{\protect\alpha \protect\beta }^{0}$ is shown here explicitly as the dashed line. The
energies $E_b$ (solid lines) are labeled by their high field quantum
numbers $|SM_{S}IM_{I}\rangle $. The binding energies of the least bound states in the singlet and triplet potentials 
are chosen to be $\protect\epsilon ^{0}=-10$ 
and $\protect\epsilon ^{1}=-5$. The avoided
crossing around $B=0$ is proportional to the hyperfine interaction $a_{hf}$ and
those between the singlet and triplet levels to the wavefunction overlap $\protect\eta _{01}$ and the hyperfine interaction $a_{hf}$. 
Four Feshbach resonances occur indicated at the
crossings $1$, $2$ and $3$ (double resonance). } \label{fig:explot}
\end{figure}

To demonstrate the procedure of identification of Feshbach
resonances we show in Fig.\thinspace \ref{fig:explot} the ABM
solutions for a simple fictitious
homonuclear system with $S=1$ and $I=2$ for an entrance channel with $M_{F}=M_{S}+M_{I}=0$ and $l=0$. 
The example has the spin structure
of $^{6}$Li but we use ABM parameters, $\epsilon ^{0}$, $\epsilon
^{1}$ and $\eta ^{01}$, with values chosen for convenience of
illustration. The field dependence of threshold energy of the
entrance channel $E_{\alpha \beta }^{0} $ is shown here explicitly
(dashed line). The energies $E_b$ (solid lines) are labeled by their high field quantum 
numbers $|SM_{S}IM_{I}\rangle $ and the binding energies in the singlet and triplet potentials are chosen 
to be $\epsilon ^{0}=-10$ and $\epsilon ^{1}=-5$.
The avoided crossings
around $B=0$ are caused by the hyperfine interaction and is
proportional to $a_{\mathrm{hf}}$; the avoided crossing between
the singlet and triplet levels is proportional to the wavefunction
overlap $\eta^{01}$. Four $s$-wave Feshbach resonances occur
indicated at
the crossings $1$, $2$ and $3$ (double resonance). The resonances at $1$ and $2$ 
are mostly determined by the triplet binding energy
$\epsilon^1$ and the resonances at $3$ by the singlet binding energy $\epsilon^0$.

\subsection{Free parameters\label{sect:vLevels}}

The free parameters of the ABM are the binding energies $\epsilon _{\nu}^{S
,l}$ and the Franck-Condon factors $\eta _{\nu ,\nu^{\prime }}^{S,S
^{\prime }}$. These parameters can be obtained in a variety of manners. Here
we discuss three methods, two of which will be demonstrated in Sect. \ref{sect:Application} and the third in Ref. \cite{heliumABM}.

First, if the scattering potentials $V_{S}^{l}(r)$ are very well known, the
bound state wavefunctions of the vibrational levels can be obtained by
solving equation (\ref{eq:Hrel1D}) for $\epsilon _{\nu}^{Sl}<0$. Numerical
values of the Franck-Condon factors follow from the obtained eigenfunctions.
This method is very accurate and can be extended to deeply bound levels,
however accurate model potentials are only available for a limited number of
systems.

A second method can be used when the potentials are not very well or only
partially known. For large interatomic distances the potentials can be
parameterized by the dispersion potential
\begin{equation}
V(r)=-\frac{C_{6}}{r^{6}}.  \label{eq:VanderWaals}
\end{equation}Since this expression is not correct for short distances, we account for the
inaccurate inner part of the potential by a boundary condition based on the
accumulated phase method \cite{accphase}. This boundary condition has a
one-to-one relationship to the interspecies $s$-wave singlet and triplet
scattering lengths. This approach requires only three input parameters: the
van der Waals $C_{6}$ coefficient and the singlet ($a_{S}$) and triplet ($a_{T}$) scattering lengths. For an accurate description involving
deeper bound states the accumulated phase boundary condition can
be made more accurate by including additional parameters
\cite{accphase}.

The third method to obtain the free parameters is by direct
comparison of ABM predictions with experimentally observed
Feshbach resonances, for instance obtained in a search for loss
features in an ultracold atomic gas. A loss feature spectrum can
be obtained by measuring, as a function of magnetic field, the
remaining number of atoms after a certain holding time at fixed
magnetic field. The ABM parameters follow by a fitting procedure
yielding the best match of the predicted threshold crossing fields
with the observed loss feature spectrum. We applied this method in
Ref.~\cite{wille}, where it has proven to be very powerful for
rapid assignment of Feshbach resonances in the $^6$Li-$^{40}$K mixture due to the small
computational time required to diagonalize a spin hamiltonian.

The number of fit parameters is determined by the number of bound states
which have to be considered. Depending on the atomic species and the
magnetic field, only a selected number of vibrational levels $\epsilon _{\nu}^{Sl}$ 
have to be taken into account. This number can be estimated by
considering the maximum energy range involved. An upper bound results from
comparing the maximum dissociation energy of the least bound vibrational
level $D^{\ast }$ with the maximum internal energy of the atom pair $E_{\mathrm{int,max}}$. 
The maximum dissociation energy of the $\nu$-th vibrational level 
can be estimated semi-classically
\cite{LeRoy70}
\begin{equation}
D^{\ast }=\left( \frac{\nu \zeta \hbar }{\mu ^{1/2}C_{6}^{1/6}}\right) ^{3}
\label{eq:Dstar}
\end{equation}where $\zeta =2\left[ \Gamma (1+1/6)/\Gamma (1/2+1/6)\right]
\simeq 3.434$ where $\nu$ is counted from the
dissociation limit, i.e. $\nu=1$ is the least bound state. The
maximum internal energy is given by the sum of the hyperfine
splitting of each of the two atoms at zero field, the maximum
Zeeman energy for the free atom pair and the maximum Zeeman energy
for the molecule
\begin{equation}
\label{eq:Eintmax}
E_{\mathrm{int,max}}\simeq E_{hf}^{\alpha }+E_{hf}^{\beta }+2(s_{\alpha
}+s_{\beta })g_{S}\mu _{B} B,
\end{equation}
where $E_{hf}^{\alpha ,\beta }=a_{hf}^{\alpha ,\beta }(i_{\alpha ,\beta
}+s_{\alpha ,\beta })$ and we have neglected the nuclear Zeeman effect.
Comparing equations (\ref{eq:Dstar}) and (\ref{eq:Eintmax}) gives us an expression for the number of
vibrational levels $N_{\nu}$ which have to be considered
\begin{equation}
N_{\nu}\simeq \lceil \frac{\mu ^{1/2}C_{6}^{1/6}}{\hbar \zeta }E_{\mathrm{int,max}}^{1/3}\rceil  \label{eq:Nstates}
\end{equation}where $\lceil x\rceil $ denotes the smallest integer not less than
the argument $x$. The maximum possible magnetic field $B_{max}$
required to encounter a Feshbach resonance can be estimated from
Eq. \ref{eq:Dstar} by neglecting the hyperfine energy as
\begin{equation}
B_{max} \simeq \frac{D^{\ast }}{(s_{\alpha }+s_{\beta })g_{S}\mu _{B}}.
\end{equation}
If the hyperfine energy is comparable or larger than the
vibrational level splitting $D^{\ast }$ the expression for
$B_{max}$ overestimates the maximum field of the lowest field
Feshbach resonance.

\subsection{Asymptotic bound states\label{sect:ABS}}

The most crucial ABM parameters are the binding energies $\epsilon _{\nu}^{Sl}$. 
However, for accurate predictions of the Feshbach resonance positions
also the Franck-Condon factors have to be accurate. For weakly bound states
these factors are mainly determined by the difference in binding energy of
the overlapping states, rather than by the potential shape. Therefore good
approximations can be made with little knowledge of the scattering potential.

For very-weakly bound states the outer classical turning point $r_{c}$ is found at
distances $r_{c}\gg r_{0}$; i.e., far beyond the Van der Waals radius of the
interaction potential
\begin{equation}
r_{0}=\frac{1}{2}\left( \frac{2\mu C_{6}}{\hbar ^{2}}\right) ^{1/4}.  \label{eq:rvdW}
\end{equation}These states are called \emph{halo states}\cite{haloref}. Because in this
case most of the probability density of the wavefunction is found outside
the outer classical turning point, these states can be described by a
zero-range potential with a wavefunction of the type $\psi (r)\sim
e^{-\kappa r}$, where $\kappa =(-2\mu \epsilon /\hbar ^{2})^{1/2}$ is the
wavevector corresponding to a bound state with binding energy $\epsilon $.
The Franck-Condon factor of two halo states with wavevectors $\kappa _{0}$
and $\kappa _{1}$ is given by
\begin{equation}
\langle \psi ^{0}|\psi ^{1}\rangle =2\frac{\sqrt{\kappa _{0}\kappa _{1}}}{\kappa _{0}+\kappa _{1}}.  \label{eq:overlapContact}
\end{equation}
This approximation is valid for binding energies $|\epsilon |\ll C_{6}/r_{0}^{6}$.

The calculation of the Franck-Condon factors can be extended to deeper bound
states by including the dispersive van der Waals tail. For distances $r\gg
r_{X}$, where $r_{X}$ is the exchange radius, the potential is well
described by Eq. (\ref{eq:VanderWaals}) and the Franck-Condon factors can be calculated
by numerically solving the Schr\"{o}dinger equation (\ref{eq:Hrel1D}) for
the Van der Waals potential (\ref{eq:VanderWaals}) on the interval $r_{X}<r<\infty $ \cite{Gao98}. The exchange radius $r_{X}$ is defined as the
distance where the Van der Waals interaction equals the exchange
interaction. This method can be used for \emph{asymptotic bound states},
which we define by the condition $r_{c}>r_{X}$. If even deeper bound
states, with $r_{c}<r_{X}$, have to be taken into account, the potential can
be extended by including the exchange interaction \cite{smirnov65}, or by
using full model potentials.

\begin{figure}[ht!]
\includegraphics[width=8.3088cm]{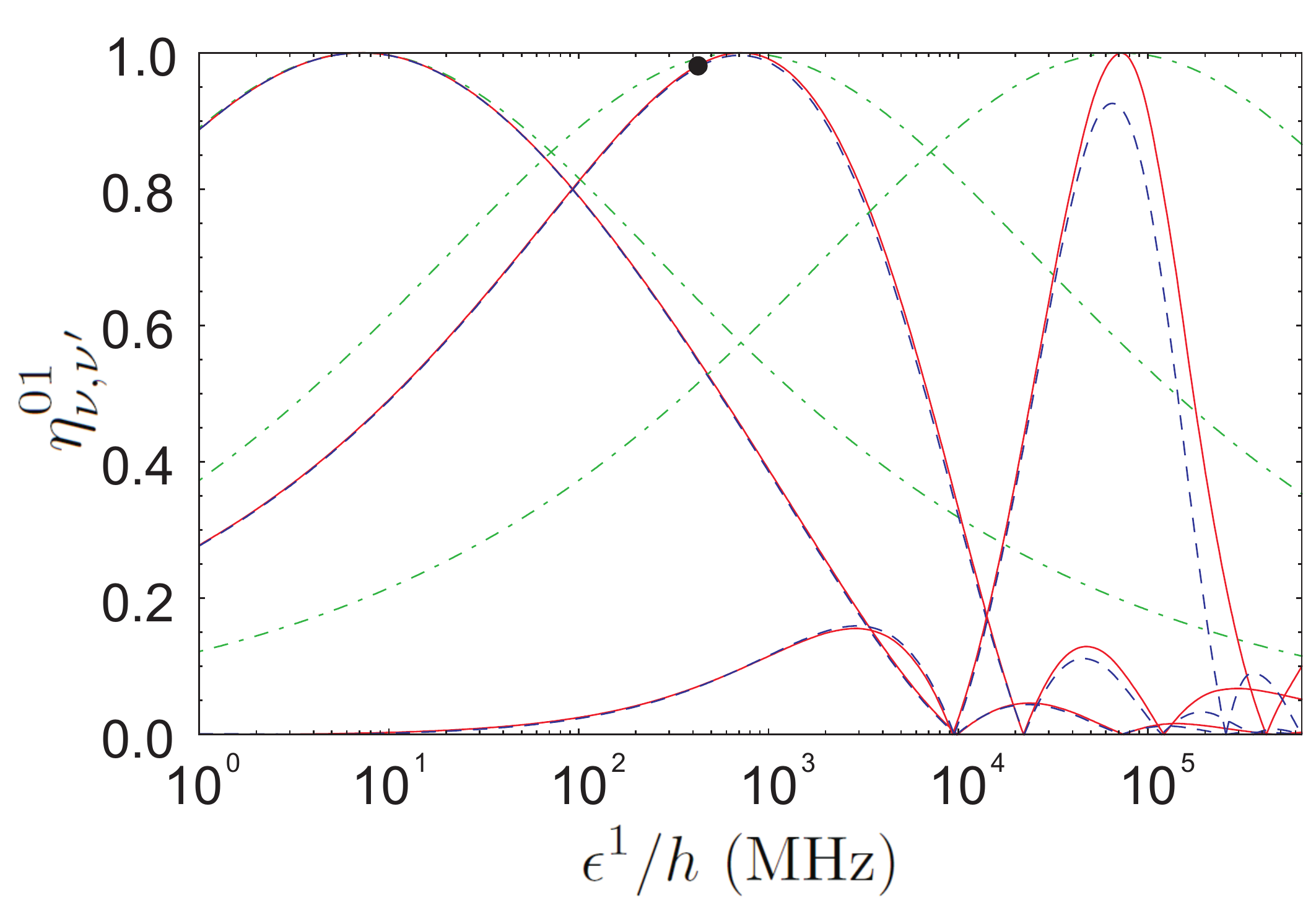}
\caption{(Color online) The Franck-Condon factor $\eta^{01}_{\nu,\nu'}$ for the
least bound states of the $^6$Li-$^{40}$K system, calculated as a function of the triplet
binding energy $\epsilon^1$ for three different values of
$\epsilon^0/h=7.16$ MHz, $\epsilon^0/h=716$ MHz and
$\epsilon^0/h=7.16\times 10^4$ MHz. $\eta^{01}_{\nu,\nu'}$ is calculated
for the model potential (dashed blue), the $-C_6/r^6$ potential (solid
red) and the contact potential, equation (\ref{eq:overlapContact})
(dash-dotted green). The black circle indicates the actual value
for the least bound state of $^{6}$Li-$^{40}$K ($\protect\epsilon ^{0}/h=716$ MHz 
and $\protect\epsilon ^{1}/h=425$ MHz). The
nodes in $\protect\eta^{01}_{\nu,\nu'}$ correspond approximately to the appearance of deeper lying vibrational
states, i.e. for $\epsilon^1/h \gtrsim 10^4~\mathrm{MHz}$, $\nu>1$.} \label{fig:overlap}
\end{figure}

To illustrate the high degree of accuracy achieved by using asymptotic
bound states we calculate the Franck-Condon factor for a contact potential
(halo states), a van der Waals potential (asymptotic bound states) and a
full model potential including short range behavior, derived
from Refs. \cite{salami07,aymar05}.
Figure \ref{fig:overlap}
shows the Franck-Condon factor $\eta _{11}^{01}$ for $^{6}$Li-$^{40}$K
calculated numerically for the model potential, van der Waals potential, and
analytically using equation (\ref{eq:overlapContact}). The van der Waals
coefficient used is $C_{6}=2322E_{h}a_{0}^{6}$ \cite{derevianko}, where $E_{h}=4.35974\times
10^{-18}~\mathrm{J}$ and $a_{0}=0.05291772~\mathrm{nm}$. The
value of $\eta _{11}^{01}$ has been plotted as a function of the triplet binding
energy $\epsilon ^{1}$ for three different values of the singlet binding
energy $\epsilon ^{0}$. It is clear that the contact potential is only
applicable for $\epsilon /h\lesssim 100$MHz, hence only for systems with
resonant scattering in the singlet and triplet channels. 

The approximation
based on the $C_{6}$ potential yields good agreement down to binding
energies of $\epsilon/h \lesssim 40$GHz, which is much more than the
maximum possible vibrational level splitting of the least bound states ($D^{\ast }/h= 8.2$GHz),
hence is sufficient to describe Feshbach resonances originating for the least bound vibrational level.
The black circle indicates the actual Franck-Condon factor for the least bound state of $^{6}$Li-$^{40}$K. 
For the contact, van der Waals and model potentials we find $\eta _{11}^{01}=0.991$,
$\eta _{11}^{01}=0.981$ and $\eta _{11}^{01}=0.979$ respectively.

\begin{figure}[ht!]
\includegraphics[width=8.3066cm]{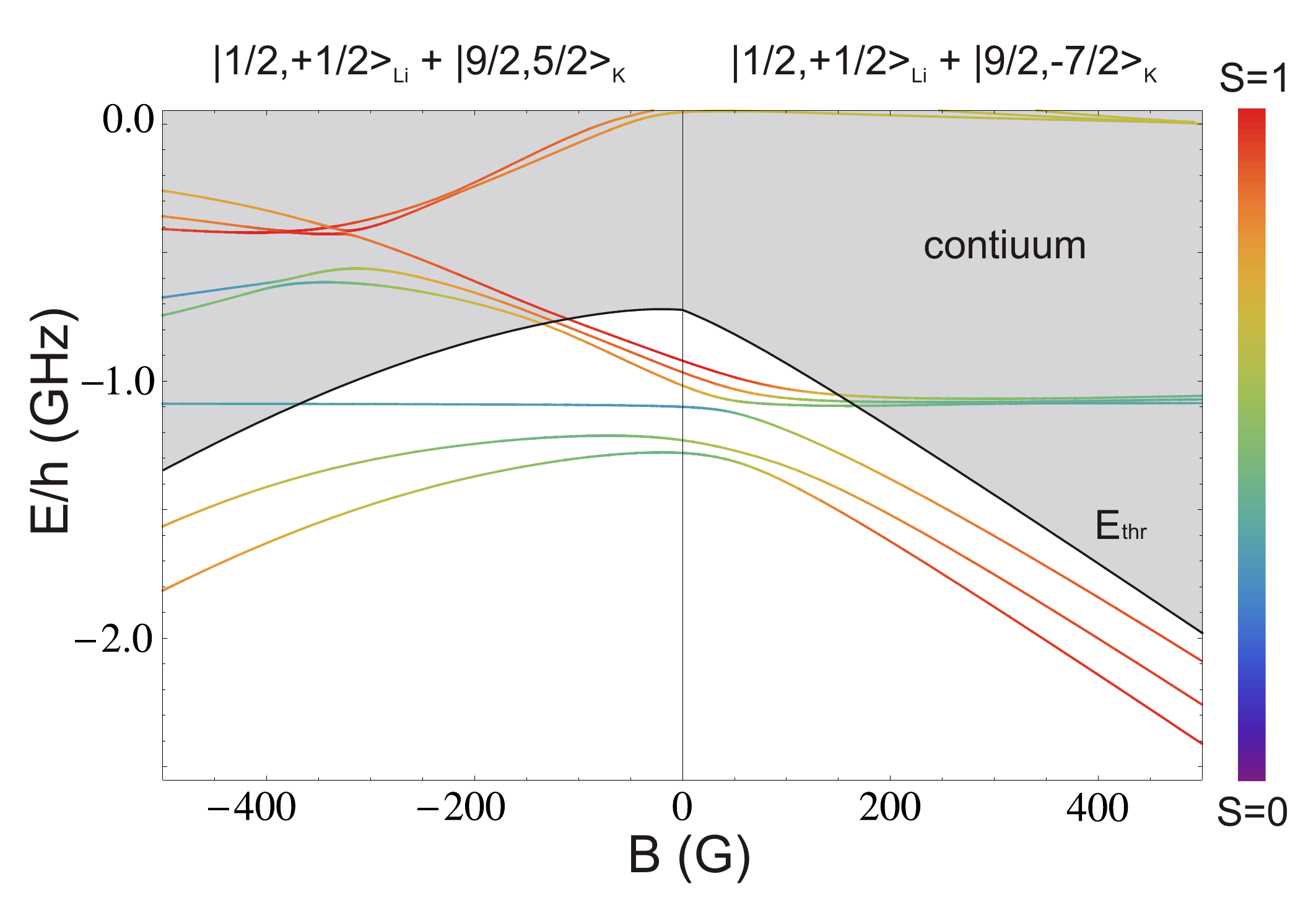}
\caption{(Color online) The energies of all the coupled bound states for $^{6}$Li -$^{40}$K with
total spin $M_{F}=\pm 3$. The black solid line indicates the
threshold energy of the entrance channel
$|1/2,+1/2\rangle _{\mathrm{Li}}\otimes |9/2,+5/2\rangle _{\mathrm{K}}$
for $B<0$ and $|1/2,+1/2\rangle _{\mathrm{Li}}\otimes |9/2,-7/2\rangle _{\mathrm{K}}$
for $B>0$. The grey area
represents the scattering continuum and the (colored) lines indicate the
coupled bound states. Feshbach resonances occur when a bound state crosses
the threshold energy. The color scheme indicates the admixture of singlet
and triplet contributions in the bound states obtained from the eigenstates of 
the Hamiltonian (\ref{1}). The strong admixture near the
threshold crossings at $B\simeq 150$ $\mathrm{G}$ demonstrate the importance
of the singlet-triplet mixing in describing Feshbach resonance positions
accurately. Since in these calculations the coupled bound states are not coupled to the
open channel, they exist even for energies above the threshold.} \label{fig:LiKSpectrum}
\end{figure}

\section{Application to various systems\label{sect:Application}}

In this section we demonstrate the versatility of the ABM by applying it to
two different systems using the different approaches as discussed in
Section \ref{sect:vLevels}.

\subsection{$^{6}$Li -$^{40}$K}

Both $^{6}$Li and $^{40}$K have electron spin $s=1/2$, therefore the total
electron spin can be singlet $S=0$ or triplet $S=1$. We intend to describe
all loss features observed in Ref.\thinspace \cite{wille}. Since all these
features were observed for magnetic fields below $300~$G we find that, by
use of Eq.~(\ref{eq:Nstates}), it is sufficient to take into account only
the least bound state ($\nu =1$) of the singlet and triplet potentials. This
reduces the number of fit parameters to $\epsilon _{\nu}^{Sl}=\epsilon
_{1}^{0,l}$ and $\epsilon _{1}^{1,l}$. Subsequently, we calculate the
rotational shifts by parameterizing the $l>0$ bound state energies with the
aid of model potentials \footnote{Note that this procedure can also be applied with only a $C_{6}$ coefficient
by utilizing the accumulated phase method.} as described by \cite{salami07,aymar05} . This allows us to reduce the
number of binding energies to be considered to only two: $\epsilon
_{1}^{0,0}\equiv \epsilon ^{0}$ and $\epsilon _{1}^{1,0}\equiv \epsilon ^{1}$. 
We now turn to the Franck-Condon factor $\eta _{11}^{01}$ of the two bound
states. As discussed in Section \ref{sect:ABS} its value is $\eta_{11}^{01}=0.979$ 
and can be taken along in the calculation or approximated
as unity. We first consider the case of $\eta _{11}^{01}\equiv 1$, this
reduces the total number of fit parameters to only two. We fit the positions
of the threshold crossings to the 13 observed loss features reported in
Ref.\thinspace \cite{wille} by minimizing the $\chi ^{2}$ value while
varying $\epsilon ^{0}$ and $\epsilon ^{1}$. We obtain optimal values of 
$\epsilon ^{0}/h=716(15)~\mathrm{MHz}$ and $\epsilon ^{1}/h=425(5)~\mathrm{MHz}$, 
where the error bars indicate one standard deviation. In Fig.\thinspace \ref{fig:LiKSpectrum}, 
the threshold and spectrum of coupled bound states
with $M_{F}=+3(-3)$ is shown for positive (negative) magnetic field values.
The color scheme indicates the admixture of singlet and triplet
contributions in the bound states. Feshbach resonances will occur at
magnetic fields where the energy of the coupled bound states and the
scattering threshold match. The strong admixture of singlet and triplet
contributions at the threshold crossings emphasizes the importance of
including the singlet-triplet mixing term $\mathcal{H}_{\mathrm{hf}}^{-}$
in the Hamiltonian. All 13 calculated resonance positions have good
agreement with the coupled channel calculations as described in
Ref.\thinspace \cite{wille}, verifying that the ABM yields a good
description of the threshold behavior of the $^{6}$Li$-^{40}$K system for
the studied field values.

We repeat the $\chi^2$ fitting procedure now including the numerical value
of the overlap. The value of $\eta^{01}_{11}$ for both the $s$-wave and $p$-wave
bound states are calculated numerically while varying $\epsilon^0$ and $\epsilon^1$. 
This fit results in a slightly larger $\chi^2$ value with
corresponding increased discrepancies in the resonance positions. However,
all of the calculated resonance positions are within the experimental widths
of the loss features. Therefore, the analysis with $\eta^{01}_{11}\equiv 1$
and $\eta^{01}_{11}=0.979$ can be safely considered to yield the same
results within the experimental accuracy.

\subsection{$^{40}$K -$^{87}$Rb}

We now turn to the $^{40}$K -$^{87}$Rb mixture to demonstrate the
application of the ABM to a system including multiple (three) vibrational
levels in each potential and the corresponding non-trivial values for the 
Franck-Condon factors. We consider $s$-wave ($l=0$) resonances. Although accurate
K-Rb scattering potentials are known \cite{PashovKRb}, we choose to use the
accumulated phase method as discussed in Section \ref{sect:vLevels} using
only three ABM parameters to demonstrate the accuracy of the ABM for a
more complex system like $^{40}$K -$^{87}$Rb.

\begin{figure}[ht!]
\includegraphics[width=8.3088cm]{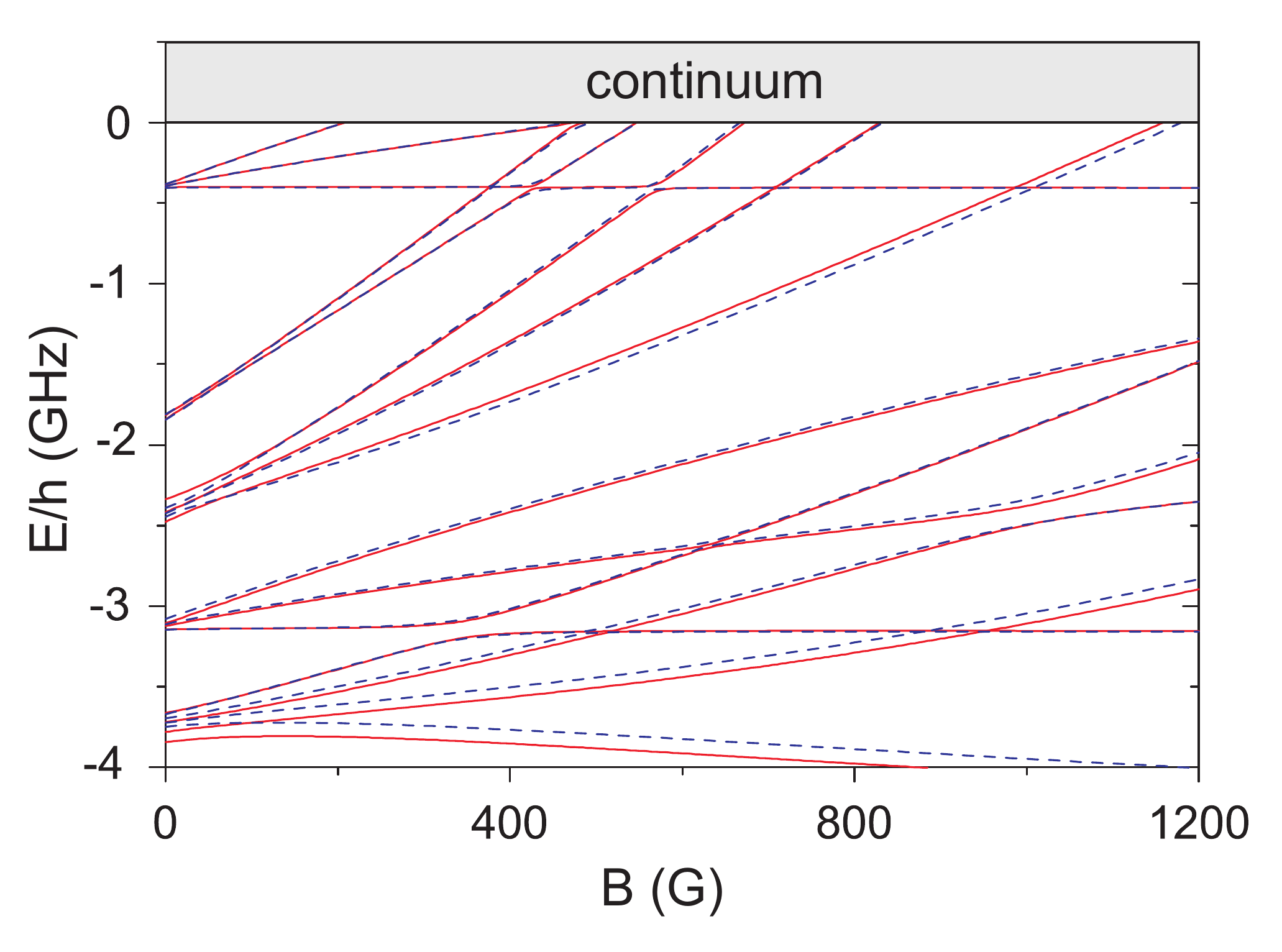}
\caption{(Color online) The bound state
spectrum for $^{40}$K -$^{87}$Rb for $M_{F}=-7/2$ plotted with respect to
the threshold energy $E_{\mathrm{K},\mathrm{Rb}}^{0}$ of 
the $|9/2,-9/2\rangle _{\mathrm{K}}$ + $|1,1\rangle _{\mathrm{Rb}}$
mixture. Solid red lines are ABM
calculations and the blue dashed lines are numerical coupled channels
calculations. Good agreement between the two calculations is found in
particular for the weakest bound levels.} \label{fig:KRbSpectrum}
\end{figure}

We solve the reduced radial wave equation (\ref{eq:Hrel1D}) 
for $V_{S}(r)=-C_{6}/r^{6}$ and the continuum state $E =\hbar ^{2}k^{2}/2\mu $
in the limit $k\rightarrow 0$. We obtain the accumulated phase
boundary condition at $r_{in}=18~a_{0}$ from the boundary condition at $r\rightarrow \infty $
using the asymptotic $s$-wave scattering phase shift $\delta _{0}=\arctan (-ka)$, 
where $a$ is the known singlet or triplet scattering length. Subsequently, we obtain binding energies for the
three last bound states of the singlet and triplet potential by solving the
same equation (\ref{eq:Hrel1D}) but now using the accumulated phase
at $r=r_{in}$ and $\psi (r\rightarrow \infty )=0$ as boundary conditions. We
numerically calculate the Franck-Condon factors by normalizing the
wavefunctions for $r_{in}<r<r_{\infty}$, thereby neglecting the wavefunction in the
inner part of the potential ($0<r<r_{in}$). This approximation becomes less
valid for more deeply bound states. We use as input parameters $C_{6}=4274~E_{h}a_{0}^{6}$ \cite{derevianko}, 
$a_{S}=-111.5~a_{0}$ and $a_{T}=-215.6~a_{0}$ \cite{PashovKRb}.
Figure \ref{fig:KRbSpectrum} shows the
spectrum of bound states with respect to the threshold energy for the spin
mixture of $|f,m_{f}\rangle =|9/2,-9/2\rangle _{\mathrm{K}}$ and $|1,1\rangle _{\mathrm{Rb}}$ states. 
The red curves indicate the ABM results
and the blue curves correspond to full coupled channel calculations \cite{JulienneKRb}. 
The agreement between the two models is satisfactory,
especially for the weakest bound states close to the threshold. 
A conceptually different analysis of the K-Rb system using also only three input parameters 
has been performed by Hanna, \emph{et al.} \cite{hanna09}.

\section{Feshbach Resonance Width}
\label{sect:OpenChannel}
\subsection{Overview}
The asymptotic bound state model has been
used so far to determine the position of the Feshbach resonances
but not their width. As is well known from standard Feshbach
theory, the width of $s$-wave resonances depends on the coupling
between the resonant level and the continuum \cite{fesh1,fesh2}.
For resonances with $l>0$ the width is determined by a physically
different process, namely tunneling through the centrifugal
barrier. Here we discuss only the width of $s$-wave resonances. We
determine the resonance width by analyzing the shift of the
resonant level close to threshold due to the coupling with the
least bound state of the open channel. This is possible using
again the restricted basis set of bound states $\{|\psi _{\nu}^{Sl}\rangle
|\sigma \rangle \}$, introduced
in Section \ref{sect:ABM}. 
The possibility to obtain the resonance width by analyzing the shift 
is plausible because near a resonance the scattering behavior in the zero energy limit
 is closely related to the threshold behavior of the bound-state.
To reveal the coupling as contained in the ABM
approximation we partition the total Hilbert space of the
Hamiltonian (\ref{1}) into two orthogonal subspaces $\mathcal{P}$
and $\mathcal{Q}$. The states of the open channels are located in
$\mathcal{P}$ space, those of the closed channels in $\mathcal{Q}$
space \cite{moerdijk}. This splits the Hamiltonian $\mathcal{H}$
in four parts (cf. Section \ref{sect:tailFeshbach});
$\mathcal{H} = \mathcal{H}_{PP}+\mathcal{H}_{PQ}+\mathcal{H}_{QP}+\mathcal{H}_{QQ}$.
Here
$\mathcal{H}_{PP}$ and $\mathcal{H}_{QQ}$ describe the system
within each subspace and
$\mathcal{H}_{PQ}(=\mathcal{H}_{QP}^\dag)$ describe the coupling
between the $\mathcal{P}$ and $\mathcal{Q}$ spaces, thus providing
a measure for the coupling between the open channels in
$\mathcal{P}$ space and the closed channels in $\mathcal{Q}$ space.

The scattering channels are defined by the Breit-Rabi pair states $|\alpha
\beta \rangle =|f_{\alpha }m_{f_{\alpha }},f_{\beta }m_{f_{\beta }}\rangle $. In the associated pair basis 
the diagonal matrix elements of the
Hamiltonian $\mathcal{H}$ correspond to the 'bare' binding energies of the 
pair states; i.e. the binding energy in the abscence of coupling between the channels by $V(r)$.
Restricting ourselves, for
purposes of introduction, to a single open channel and to the least bound states in the interaction
potentials, $\mathcal{H}_{PP}$ is a single matrix element on the diagonal of $\mathcal{H}$, corresponding to the
\textit{bare} binding energy of the least bound state of the entrance channel, $\epsilon _{P}=-\hbar
^{2}\kappa _{P}^{2}/2\mu $. The energy $\epsilon _{P}$ can be readily
calculated by projecting the pair state on the spin basis 
$\{|\sigma \rangle =|SM_{s}\mu _{\alpha }\mu _{\beta }\rangle \}$, and is
given by

\begin{equation}
\epsilon _{P}=\sum_{S}\epsilon _{\nu}^{Sl}\!\!\!\sum_{M_{S},\mu _{\alpha
},\mu _{\beta }}\langle SM_{s}\mu _{\alpha }\mu _{\beta }|f_{\alpha
}m_{f_{\alpha }},f_{\beta }m_{f_{\beta }}\rangle ^{2}
\label{eq:epsilonPDIS2}
\end{equation}In the following sections we will show that the width $\Delta B$ of the
resonance is a function of the bare binding energy $\epsilon _{P}$ of
the entrance channel and a single matrix element of $\mathcal{H}_{PQ}$, denoted
by $\mathcal{K}$, representing the coupling of the level $\epsilon _{P}$
to the resonant level in $\mathcal{Q}$ space. We will show that the width is
given by the expression
\begin{equation}
\mu _{rel}\Delta B=\frac{\mathcal{K}^{2}}{2a_{bg}\kappa _{P}|\epsilon _{P}|}.
\end{equation}
Hence, once the ABM parameters are known, the width of the
resonances follows with an additional unitary transformation of the ABM
matrix to obtain the coupling coefficient $\mathcal{K}$. In view of the
orthogonality of the subspaces $\mathcal{P}$ and $\mathcal{Q}$, the
submatrix $\mathcal{H}_{QQ}$, corresponding to all closed channels, can be
diagonalized, leaving $\mathcal{H}_{PP}$ unaffected but changing 
the $\mathcal{H}_{PQ}$ and $\mathcal{H}_{QP}$ submatrices. In diagonalized form
the $\mathcal{H}_{QQ}$ submatrix contains the energies $\epsilon _{Q}$ of
all bound levels in $\mathcal{Q}$ space and includes the coupling of all
channels except the coupling to $\mathcal{P}$ space. This transformation
allows to identify the resonant bound state and the corresponding
off-diagonal matrix element $\mathcal{K}$ in $\mathcal{H}_{PQ}$, which is a
measure for the resonance width. Thus we can obtain the coupling between the
open and closed channels without the actual use of continuum states.

In Section \ref{sect:tailFeshbach} we present the Feshbach theory tailored
to suit the ABM. We give a detailed description of the resonant coupling,
and demonstrate with a two-channel model how the ABM bound state energy $E_b$ compares to
the associated $\mathcal{P}$-space bare energy $\epsilon_{P}$, and to the dressed level in 
the entrance channel from which one can deduce the resonance width. 
In Section \ref{sect:dressABM} we generalize the results such that the width of the
Feshbach resonances can be obtained for the general multi-channel case. For
a more thorough treatment of the Feshbach formalism we refer the reader to
\cite{fesh1,fesh2} and for its application to cold atom scattering e.g. \cite{moerdijk}.

\subsection{Tailored Feshbach theory}

\label{sect:tailFeshbach}By introducing the projector operators $P$ and $Q$,
which project onto the subspaces $\mathcal{P}$ and $\mathcal{Q}$,
respectively, the two-body Schr\"{o}dinger equation can be split into a set
of coupled equations \cite{moerdijk}
\begin{eqnarray}
(E-\mathcal{H}_{PP})|\Psi _{P}\rangle &=&\mathcal{H}_{PQ}|\Psi _{Q}\rangle ,
\label{eq:P} \\
(E-\mathcal{H}_{QQ})|\Psi _{Q}\rangle &=&\mathcal{H}_{QP}|\Psi _{P}\rangle ,
\label{eq:Q}
\end{eqnarray}where $|\Psi _{P}\rangle \equiv P|\Psi \rangle $, $|\Psi _{Q}\rangle \equiv
Q|\Psi \rangle $, $\mathcal{H}_{PP}\equiv P\mathcal{H}P$, $\mathcal{H}_{PQ}\equiv P\mathcal{H}Q$, etc. Within $\mathcal{Q}$ space the Hamiltonian $\mathcal{H}_{QQ}$ is diagonal with eigenstates $|\phi _{Q}\rangle $
corresponding to the two-body bound state with eigenvalues $\epsilon _{Q}$.
The energy $E=\hbar ^{2}k^{2}/2\mu $ is defined with respect to the open
channel dissociation threshold.

We consider one open channel and assume that near a resonance it couples to
a single closed channel. This allows us to write the $S$ matrix of the
effective problem in $\mathcal{P}$ space as \cite{moerdijk}
\begin{equation}  \label{eQ:SmatrixSRA}
S(k)=S_P(k)\Bigg(1-2\pi i\frac{\vert\langle\phi_Q\vert \mathcal{H}_{QP}\vert\Psi_P^+\rangle\vert^2}{E-\epsilon_Q-\mathcal{A}(E)}\Bigg),
\end{equation}
where $|\Psi_P^+\rangle$ are scattering eigenstates of $\mathcal{H}_{PP}$, $S_P(k)$ is the direct scattering matrix describing the scattering process in
$\mathcal{P}$ space in the absence of coupling to $\mathcal{Q}$ space.

The complex energy shift $\mathcal{A}(E)$ describes the dressing of the bare
bound state $|\phi_Q\rangle$ by the coupling to the $\mathcal{P}$ space and
is represented by
\begin{equation}  \label{eQ:AEcomplex}
\mathcal{A}(E)=\langle \phi_Q\vert \mathcal{H}_{QP}\frac{1}{E^+-\mathcal{H}_{PP}}\mathcal{H}_{PQ}\vert \phi_Q\rangle,
\end{equation}
where $E^+=E+i\delta$ with $\delta$ approaching zero from positive values.
Usually the open channel propagator $[E^+-\mathcal{H}_{PP}]^{-1}$ is
expanded to a complete set of eigenstates of $\mathcal{H}_{PP}$, where the
dominant contribution comes from scattering states. To circumvent the use of
scattering states we expand the propagator to Gamow resonance states, which
leads to a Mittag-Leffler expansion \cite{bout}
\begin{equation}  \label{eq:Mittag}
\frac{1}{E^+-\mathcal{H}_{PP}}=\frac{\mu}{\hbar^2}\sum_{n=1}^\infty\frac{|\Omega_n\rangle\langle\Omega_n^D|}{k_n(k-k_n)},
\end{equation}
where $n$ runs over all poles of the $S_P$ matrix. The Gamow state $|\Omega_n\rangle$ is an eigenstate of $\mathcal{H}_{PP}$ with eigenvalue $\epsilon_{P_n}=\hbar^2k_n^2/(2\mu)$. Correspondingly, the dual state $|\Omega_n^D\rangle \equiv |\Omega_{n}\rangle^*$, is an eigenstate of $\mathcal{H}_{PP}^\dag$ with eigenvalue $(\epsilon_{P_n})^*$. Using these
dual states, the Gamow states form a biorthogonal set such that $\langle
\Omega_n^D\vert \Omega_{n^{\prime }}\rangle=\delta_{nn^{\prime }}$. For
bound-state poles $k_n=i\kappa_n$, where $\kappa_n>0$, Gamow states
correspond to properly normalized bound states.

We assume the scattering in the open channel is dominated by a single bound
state ($k_{n}=i\kappa _{P}$). This allows us to write the direct scattering
matrix in Eq.~(\ref{eQ:SmatrixSRA}) as
\begin{equation}
S_{P}(k)=e^{-2ika_{\mathrm{bg}}}=e^{-2ika_{\mathrm{bg}}^{P}}\frac{\kappa
_{P}-ik}{\kappa _{P}+ik}
\end{equation}where $a_{\mathrm{bg}}$ is the open channel scattering length, and the $P$-channel background scattering length $a_{\mathrm{bg}}^{P}$ is on the order
of the range of the interaction potential $a_{\mathrm{bg}}^{P}\approx r_{0}$. Since we only have to consider one bound state pole (with energy $\epsilon
_{P}$) in $\mathcal{P}$ space, the Mittag-Leffler series Eq.~(\ref{eq:Mittag}) is reduced to only one term. Therefore, the complex energy shift Eq.~(\ref{eQ:AEcomplex}) reduces to
\begin{equation}
\mathcal{A}(E)=\frac{\mu }{\hbar ^{2}}\frac{-i\mathrm{A}}{\kappa
_{P}(k-i\kappa _{P})}.  \label{eq:ADR}
\end{equation}where $\mathrm{A}\equiv \langle \phi _{Q}|\mathcal{H}_{QP}|\Omega
_{P}\rangle \langle \Omega _{P}^{D}|\mathcal{H}_{PQ}|\phi _{Q}\rangle $ is a
positive constant. The coupling matrix element between open-channel bound
state and the closed-channel bound state responsible for the Feshbach
resonance is related to $\mathrm{A}$.

The complex energy shift can be decomposed into a real and imaginary part
such that $\mathcal{A}(E)=\Delta _{\mathrm{res}}(E)-\frac{i}{2}\Gamma (E)$.
For energies $E>0$ the unperturbed bound state becomes a quasi-bound state:
its energy undergoes a shift $\Delta _{\mathrm{res}}$ and acquires a finite
width $\Gamma $. For energies below the open-channel threshold, i.e. $E<0$, $\mathcal{A}(E)$ is purely real and $\Gamma (E)=0$. In the low-energy limit, $k\rightarrow 0$, Eq.~(\ref{eq:ADR}) reduces to
\begin{equation}
\mathcal{A}(E)=\Delta -iCk,
\end{equation}where $C$ is a constant characterizing the coupling strength between $\mathcal{P}$ and $\mathcal{Q}$ space \cite{moerdijk}, given by $C=\mathrm{A}(2\kappa _{P}|\epsilon _{P}|)^{-1}$ and $\Delta =\mathrm{A}(2|\epsilon
_{P}|)^{-1}$. Note that if the direct interaction is resonant, $|a_{\mathrm{bg}}|\gg r_{0}$, the energy dependence of the complex energy shift is given
by \cite{marcelis06} $\mathcal{A}(E)=\Delta -iCk(1+ika^{P})^{-1}$ where $a^{P}=\kappa _{P}^{-1}$, yielding an energy dependence of the resonance
shift.\begin{figure}[ht!]
\includegraphics[width=8.3088cm]{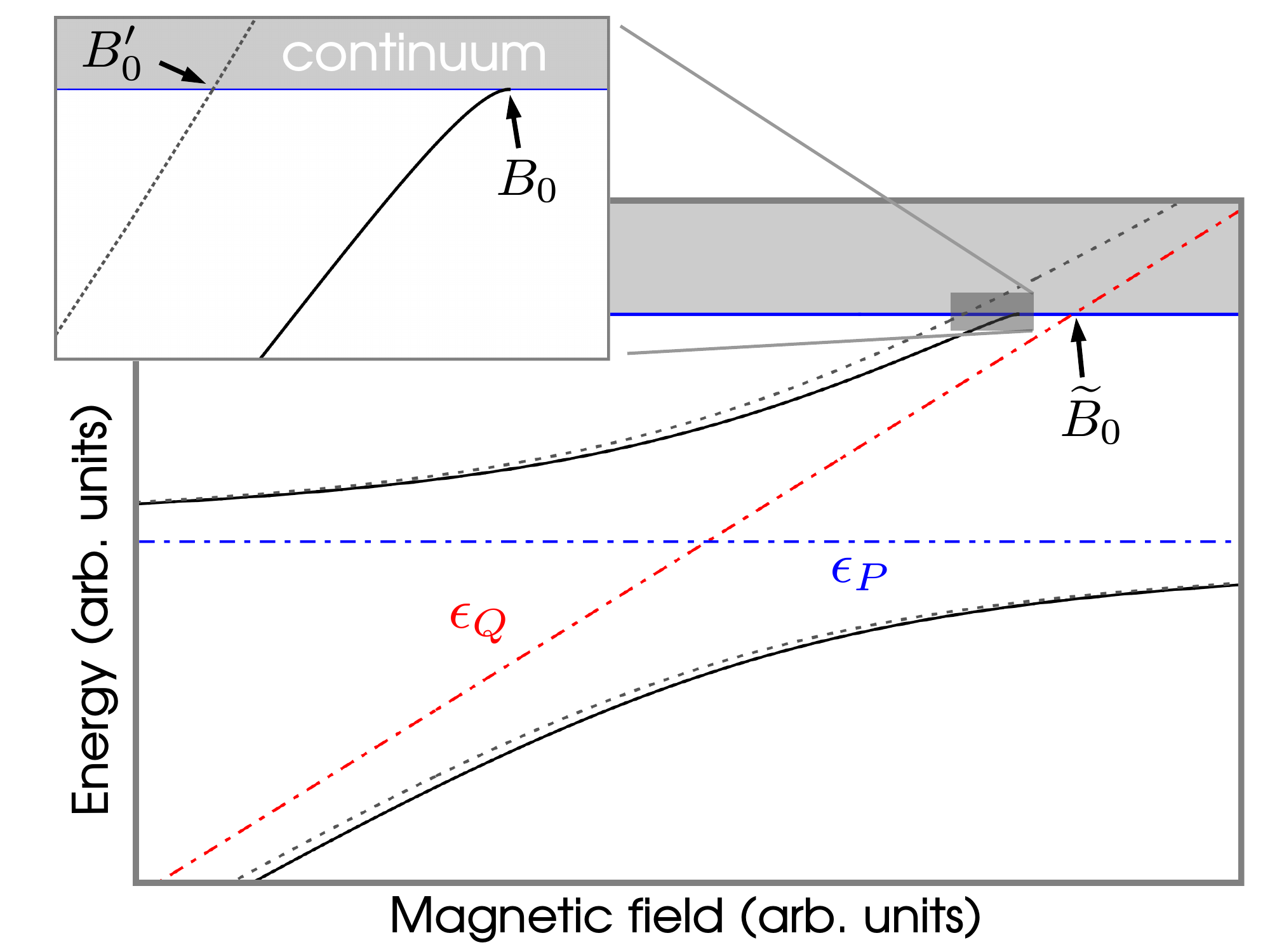}
\caption{(Color online) 
Illustration of the threshold behavior in a (fictious) two-channel version of the dressed ABM.
The threshold behavior is determined by the coupling between the least bound level in the 
open channel in $\mathcal{P}$ space with the resonant bound level in $\mathcal{Q}$ space. The uncoupled 
levels are shown as the blue ($\epsilon_P$) and red ($\epsilon_Q$) dash-dotted lines, with 
$\epsilon_Q$ crossing the threshold at $\widetilde{B}_{0}$.  The solid black lines represent the 
dressed levels, with the upper branch crossing the threshold at $B_0$. Near the 
threshold, the dressed level shows the characteristic quadratic dependence on 
($B-B_0$) (see inset). For pure ABM levels (dotted gray) no threshold effects occur 
and the coupled bound state crosses the threshold at $B_{0}^{\prime }$.}
\label{SRplot}
\end{figure}

Since we consider one open channel, the (elastic) $S$-matrix element can be
written as $e^{2i\delta (k)}$, where $\delta (k)$ is the scattering phase
shift. The scattering length, defined as the limit $a\equiv -\tan \delta
(k)/k,\,(k\rightarrow 0)$, is found to be Eq. (\ref{eq:dispersiveA})
which shows the well known dispersive behavior. The direct scattering
process is described by the scattering length $a_{\mathrm{bg}}=a_{\mathrm{bg}}^{P}+a^{P}$. 
At magnetic field value $B_{0}$, where the \emph{dressed}
bound state crosses the threshold of the entrance channel, the scattering length has a
singularity.

The dressed state can be considered as a (quasi-) bound state of the total
scattering system. The energy of these states is obtained by finding the
poles of the total $S$ matrix Eq.~(\ref{eQ:SmatrixSRA}). This results in
solving
\begin{equation}
(k-i\kappa _{P})\left( E-\epsilon _{Q}-\mathcal{A}(E)\right) =0,
\label{eq:PoleEqn}
\end{equation}for $k$. Due to the underlying assumptions, this equation is only valid for
energies around threshold where the open and closed channel poles dominate.
Near threshold, the shifted energy of the uncoupled molecular state, $\epsilon _{Q}+\Delta $, can be approximated by $\mu _{rel}(B-B_{0})$. This
allows to solve Eq. (\ref{eq:PoleEqn}) for $E$ and we readily obtain

\begin{equation}
E=-\left( \frac{2|\epsilon _{P}|^{3/2}\mu _{rel}(B-B_{0})}{\mathrm{A}}\right) ^{2}
\end{equation}
retrieving the characteristic quadratic threshold behavior of the dressed level 
as a function of ($B-B_0$). The energy dependence of molecular state close to
resonance is also given by $E=-\hbar ^{2}/(2\mu a^{2})$ this allows us to
express the field width of the resonance as $\Delta B=C(a_{\mathrm{bg}}\mu
_{rel})^{-1}$.

We apply the above Feshbach theory to a (fictitious) two-channel version of
ABM, and the results are shown in Fig.~\ref{SRplot}. This two-channel system
is represented by a symmetric $2\times 2$ Hamiltonian matrix, where there is only one
open and one closed channel. The open and closed channel binding energies $\epsilon _{P}$ resp. $\epsilon _{Q}$ are given by the diagonal matrix
elements, while the coupling is represented by the (identical) off-diagonal
matrix elements. The closed channel bound state is made linearly dependent
on the magnetic field, while the coupling is taken constant. In addition to $\epsilon _{P}$ and $\epsilon _{Q}$, we plot the corresponding ABM solution,
which in this case is equivalent to a typical two-level avoided crossing
solution. The figure now nicely illustrates the evolution from ABM to the
dressed ABM approach, where the latter solutions are found from the two
physical solutions of Eq.~(\ref{eq:PoleEqn}), which are also plotted. Since
the dressed ABM solutions account for threshold effects, they show the
characteristic quadratic bending towards threshold as a function of magnetic
field. From this curvature the resonance width can be deduced.

\begin{figure}[ht!]
\includegraphics[width=8.3088cm]{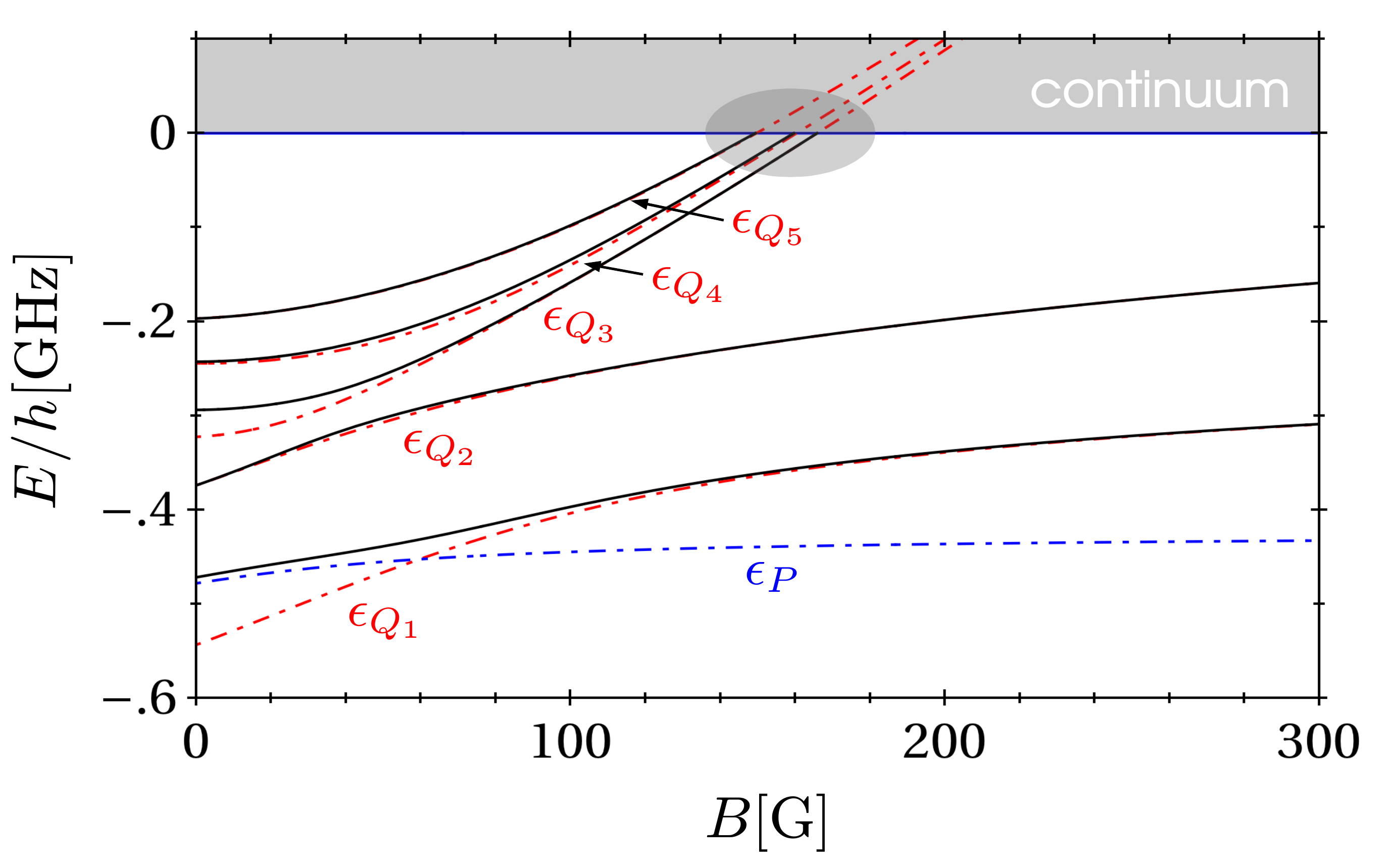}
\caption{(Color online) Dressed
bound states for $^6$Li-$^{40}$K for $M_{F}=-3$ (black lines, see also Table
\protect\ref{tab:ABMplusresult}). The uncoupled $\mathcal{Q}$ and $\mathcal{P}$ bound 
states ($\mathcal{H}_{PQ}=\mathcal{H}_{QP}=0$) are represented by the dot-dashed
lines (red and blue respectively). The gray shaded area is shown in detail in Fig. \ref{fig:ABMplusLiKmf3scat}.}
\label{fig:ABMplusLiKmf3}
\end{figure}

\subsection{The dressed Asymptotic Bound state Model}

\label{sect:dressABM}

To illustrate the presented model for a realistic case, we will discuss the $^6$Li-$^{40}$K system prepared in the $|f_{\mathrm{Li}} m_{f_{\mathrm{Li}}},f_\mathrm{K} m_{f_\mathrm{K}}\rangle= |1/2,+1/2,9/2,-7/2\rangle$ two-body
hyperfine state as an example throughout this section. This particular
mixture is the energetically lowest spin combination of the $M_F=-3$
manifold, allowing to consider only one open channel. We note that the model
can be utilized to cases containing more open channels.

In order to calculate the width of a Feshbach resonance using the method
presented in Sect. \ref{sect:tailFeshbach} three quantities are required:
the binding energy of the open channel $\epsilon_P$, of the closed channel
responsible for the Feshbach resonance $\epsilon_Q$, and the coupling matrix element
$\mathcal{K}$ between the two channels. In the following we will
describe how to obtain these quantities from the ABM by two simple basis
transformations.

For ultracold collisions the hyperfine and Zeeman interactions determine the
threshold of the various channels and thus the partitioning of the Hilbert
space into subspaces $\mathcal{P}$ and $\mathcal{Q}$, and therefore a
natural basis for our tailored Feshbach formalism consists of the
eigenstates of $\mathcal{H}^\mathrm{int}$. Experimentally a system is
prepared in an eigenstate $|\alpha \beta \rangle$ of the internal Hamiltonian $\mathcal{H}^\mathrm{int}$, which we refer to as the entrance channel (cf. 
Section \ref{sect:ABMoverview}). Performing
a basis transformation from the ${|\sigma \rangle }$ basis
to the pair basis ${|\alpha \beta \rangle }$ allows us to identify the
open and closed channel subspace. The open channel has the same
spin-structure as the entrance channel.

We now perform a second basis transformation which diagonalizes within $\mathcal{Q}$ space without affecting $\mathcal{P}$ space. We obtain the
eigenstates of $\mathcal{H}_{QQ}$ and are able to identify the bound state
responsible for a particular Feshbach resonance. The bare bound states of $\mathcal{Q}$ space are defined as $\{|\phi _{Q_{1}}\rangle ,|\phi
_{Q_{2}}\rangle ,\ldots \}$ with binding energies $\{\epsilon
_{Q_{1}},\epsilon _{Q_{2}},\ldots \}$. For the one dimensional $\mathcal{P}$
space, which is unaltered by this transformation, the bare bound state $|\Omega _{P}\rangle $ of $\mathcal{H}_{PP}$ is readily identified with
binding energy $\epsilon _{P}$. In the basis of eigenstates of $\mathcal{H}_{PP}$ and $\mathcal{H}_{QQ}$ we easily find the coupling matrix elements
between the $i$-th $\mathcal{Q}$ space bound state and the open channel
bound state $\langle \phi _{Q_{i}}|\mathcal{H}_{QP}|\Omega _{P}\rangle $.
This gives the coupling constant $\mathrm{A_{i}}=\langle \phi _{Q_{i}}|\mathcal{H}_{QP}|\Omega _{P}\rangle \langle \Omega _{P} ^{D}|\mathcal{H}_{PQ}|\phi _{Q_{i}}\rangle =\mathcal{K}^{2}$ that determines the resonance
field $B_{0}$ by solving Eq.~(\ref{eq:PoleEqn}) at threshold,
\begin{equation}
\epsilon _{Q_{i}}\epsilon _{P}=\frac{\mathcal{K}^{2}}{2}.
\end{equation}The field width of this Feshbach resonance is proportional to the magnetic
field difference between the crossings of the dressed ($B_{0}$) and
uncoupled $\mathcal{Q}$ bound states ($\widetilde{B}_{0}$) with threshold
since
\begin{equation}
\Delta B=\frac{a^{P}}{a_{\mathrm{bg}}}(B_{0}-\widetilde{B}_{0})=\frac{1}{a_{\mathrm{bg}}}\frac{\mathcal{K}^{2}}{2\kappa _{P}|\epsilon _{P}|\mu _{rel}}.
\end{equation}\begin{figure}[ht!]
\includegraphics[width=8.3088cm]{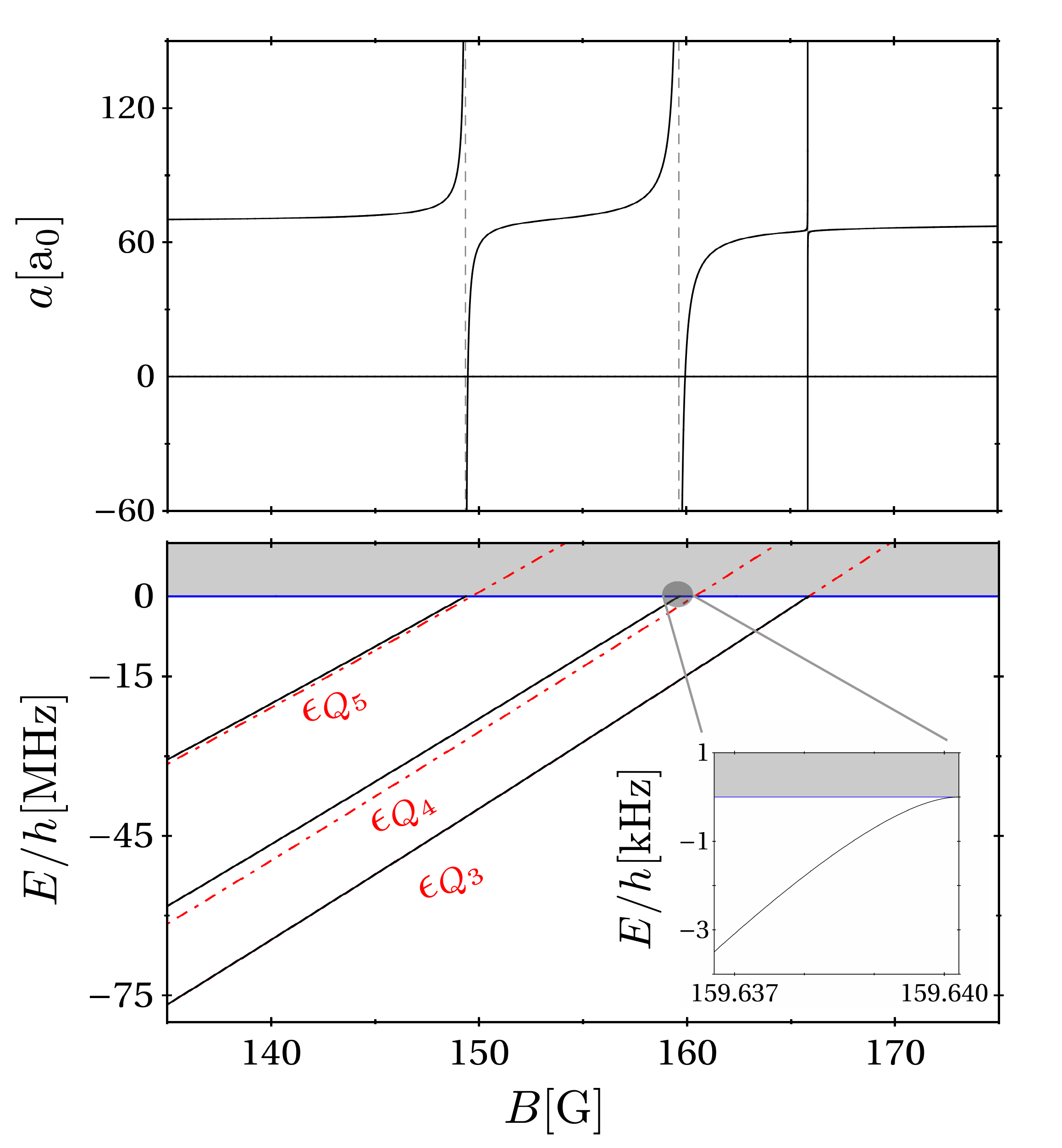}
\caption{(Color online) A zoom of the dressed ABM as shown in figure \ref{fig:ABMplusLiKmf3}. The dressed
molecular states are shown near threshold (black). The field width of a
resonance is related to the magnetic field difference of where the dressed
and uncoupled $\mathcal{Q}$ bound state cross the threshold.} \label{fig:ABMplusLiKmf3scat}
\end{figure}

We illustrate the dressed ABM for $^6$Li-$^{40}$K in Figs.~\ref{fig:ABMplusLiKmf3} and \ref{fig:ABMplusLiKmf3scat}, for $M_{F}=-3$. To demonstrate the effect of $\mathcal{H}_{PQ}$, we plotted for comparison both the uncoupled and dressed
bound states \footnote{For clarity only one of the two physical solutions is shown.}. Details of
near-threshold behavior (gray shaded area in Fig.~\ref{fig:ABMplusLiKmf3})
are shown in Fig.~\ref{fig:ABMplusLiKmf3scat} together with the obtained
scattering length. We solved the pole equation of the total $S$-matrix Eq.~(\ref{eq:PoleEqn}) for each $Q$-state and plotted only the physical solutions
which cause Feshbach resonances. The dressed bound states show the
characteristic quadratic bending near the threshold. We have used $C_{6}$ to
determine $r_{0}$ ($\approx a_{\mathrm{bg}}^{P}$) from Eq.~\ref{eq:rvdW}.

Table \ref{tab:ABMplusresult} summarizes the results of the dressed ABM for
the $^6$Li-$^{40}$K mixture. Note that the position of the Feshbach resonances will be
slightly different compared to the results from the regular ABM, for equal
values of $\epsilon_\nu^{S,0}$. Therefore, we have again preformed a $\chi^2$
analysis and we found new values of the binding energies 
$\epsilon^0/h=713~\mathrm{MHz}$ and $\epsilon^1/h=425~\mathrm{MHz}$,
which yields a lower $\chi^2$ minimum as
compared to the regular ABM calculation.

The obtained value of $\Delta B$ generally underestimates the field width of
a resonance. This originates from the fact that only the dominant bound
state pole corresponding to $a^P$ has been taken into account. By including
the pole of the dominant \emph{virtual} state in the Mittag-Leffler
expansion, the coupling between the open and closed channel will increase, 
hence, $\Delta B$ will increase.

\begin{table}
\caption{\label{tab:ABMplusresult} The positions of all experimentally
observed $s$-wave Feshbach resonances of $^6$Li$-^{40}$K. Column 2
gives the $^6$Li ($m_{f_{Li}}$) and $^{40}$K ($m_{f_K}$) hyperfine
states. For all resonances $f_{Li}=1/2$ and $f_{K}=9/2$. Note that
the experimental width of the loss feature $\Delta B_\mt{exp}$ is
not the same as the field width $\Delta B$ of the scattering
length singularity. Feshbach resonance positions $B_0$ and widths
$\Delta B$ for $^6$Li-$^{40}$K as obtained by the dressed ABM,
obtained by minimizing $\chi^2$. The last two columns show the
results of full coupled channels (CC) calculations. All magnetic
fields are given in Gauss. The experimental and CC values for
$M_F<0$ and $M_F>0$ are taken from Ref. \cite{wille} and
\cite{dePRL} respectively. The resonances marked with $^{\ast}$ have also
been studied in Refs. \cite{voigt09,spiegelhalder10}.}
\begin{tabular}{cccccccc}
\hline
\hline
& & \multicolumn{2}{c}{Experiment} & \multicolumn{2}{c}{ABM+}  & \multicolumn{2}{c}{CC}\\
$M_F$ & $m_{f_{Li}}$,$m_{f_{K}}$ & $B_0$ & $\Delta B_\mt{exp}$ & $B_0$ & $\Delta B$ & $B_0$ & $\Delta B$  \\
\hline \\ [-2ex]
-5 & $-\frac{1}{2}$, $-\frac{9}{2}$ & 215.6 & 1.7 & 216.2 & 0.16 & 215.6 & 0.25\\ [.5ex]
-4 & $+\frac{1}{2}$, $-\frac{9}{2}$ & 157.6 & 1.7 & 157.6 & 0.08 & 158.2 & 0.15 \\ [.5ex]
-4 & $+\frac{1}{2}$, $-\frac{9}{2}$ & 168.2$^{\ast}$ & 1.2 & 168.5 & 0.08 & 168.2 & 0.10 \\ [.5ex]
-3 & $+\frac{1}{2}$, $-\frac{7}{2}$ & 149.2 & 1.2 & 149.1 & 0.12 & 150.2 & 0.28 \\ [.5ex]
-3 & $+\frac{1}{2}$, $-\frac{7}{2}$ & 159.5 & 1.7 & 159.7 & 0.31 & 159.6 & 0.45 \\ [.5ex]
-3 & $+\frac{1}{2}$, $-\frac{7}{2}$ & 165.9 & 0.6 & 165.9 & 0.0002 & 165.9 & 0.001 \\ [.5ex]
-2 & $+\frac{1}{2}$, $-\frac{5}{2}$ & 141.7 & 1.4 & 141.4 & 0.12 & 143.0 & 0.36 \\ [.5ex]
-2 & $+\frac{1}{2}$, $-\frac{5}{2}$ & 154.9$^{\ast}$ & 2.0 & 154.8 & 0.50 & 155.1 & 0.81 \\ [.5ex]
-2 & $+\frac{1}{2}$, $-\frac{5}{2}$ & 162.7 & 1.7 & 162.6 & 0.07 & 162.9 & 0.60 \\ [.5ex]
+5 & $+\frac{1}{2}$, $+\frac{9}{2}$ & 114.47(5) & 1.5(5) & 115.9 & 0.91 & 114.78& 1.82 \\ [.5ex]
\hline
\hline
\end{tabular}
\end{table}


\section{Summary and Conclusion}

\label{sect:Discussion}

We have presented a model to accurately describe Feshbach resonances. 
The model allows for fast and accurate prediction of resonance
positions and widths with very little experimental input. 
The reduction of the basis to only a few states
allows to describe Feshbach resonances in a large variety 
of systems without accurate knowledge of scattering potentials. 

Using the ABM in combination with the accumulated phase method allows to describe 
Feshbach resonances in alkali systems with a large degree of accuracy, using only 
three input parameters.
Additionally, the fast computational time of the model allows to map all available Feshbach
resonances in a system and select the optimal resonance required to perform a certain experiment. 
For the $^6$Li-$^{40}$K mixture we have utilized this ability to find a broad resonance as presented
in Ref. \cite{dePRL}. In addition, 
locating e.g. overlapping resonances in multi-component (spin)mixtures can
be performed easily well using the ABM.

An additional important feature is that the model can be 
stepwise extended to include more phenomena allowing to describe more 
complex systems. For example, a possible extension would be to include the contribution of the
dominant virtual state in the Mittag-Leffler expansion, this would allow for an accurate description
of the resonance widths for systems with a large and negative $a_{\mathrm{bg}}$. Additionally, including the
dipole-dipole interaction allows to describe systems where Feshbach resonances
occur due to dipole-dipole coupling. Finally, it has already been shown by Tsherbul, et al. \cite{tscherbul10}
that the ABM can be succesfully extended by including coupling to bound states by means of
an externally applied radio-frequency field.

The approach as described in the ABM is in principle not limited to two-body systems. Magnetic field induced 
resonances in e.g. dimer-dimer scattering have already been experimentally observed \cite{chin05}. 
For few-body systems an approach without having to solve the coupled radial Schr\"odinger equations is 
very favorable. This large variety of unexplored features illustrate the richness of the model.

This work is part of
the research program on Quantum Gases of the Stichting voor Fundamenteel
Onderzoek der Materie (FOM), which is financially supported by the Nederlandse
Organisatie voor Wetenschappelijk Onderzoek (NWO).

\bibliographystyle{apsrev}
\bibliography{bibABM}


\end{document}